## Title

Experimental determination of ferric iron partitioning between pyroxene and melt at 100 KPa

## Authors

Avishek Rudra*

## Affiliation

Department of Earth and Environmental Sciences

150 John T. Tate Hall

Minneapolis

Minnesota, 55455

USA

Email id: rudra009@umn.edu

*corresponding author

Marc M. Hirschmann

## Affiliation

Department of Earth and Environmental Sciences

150 John T. Tate Hall

Minneapolis

Minnesota, 55455

USA

Email id: mmh@umn.edu

ORCID id: 0000-0003-1213-6645

## Acknowledgments

AR thanks Jed Mosenfelder and Amanda Dillman for assistance with gas mixing furnace experiments; Anette von der Handt for helping with electron microprobe analysis; Nick Seaton for assistance with EBSD data collection and Zachary Michels for helping with processing the EBSD data. AR is also thankful to Tony Lanzirotti and Matt Newville for assistance with beamline data collection and processing. AR gratefully thanks Alan Woodland and Dante Canil for providing pyroxene samples to be used as XANES standards. AR is grateful to Peat Solheid for collecting the Mössbauer spectra, which is a facility at the Institute if Rock Magnetism supported by grants from the Instrument and Facilities Program, Division of Earth Science, National Science Foundation. We gratefully acknowledge support from National Science Foundation grants EAR1426772 and EAR2016215. The XANES data collection was performed at GeoSoilEnviroCARS (The University of Chicago, Sector 13), Advanced Photon Source (APS), Argonne National Laboratory. GeoSoilEnviroCARS is supported by the National Science Foundation – Earth Sciences (EAR – 1634415) and Department of Energy- GeoSciences (DE-FG02-94ER14466). This research used resources of the Advanced Photon Source, a U.S. Department of Energy (DOE) Office of Science User Facility operated for the DOE Office of Science by Argonne National Laboratory under Contract No. DE-AC02-06CH11357.

## Abstract

Pyroxene is the principal host of $Fe^{3+}$ in basalt source regions, hosting 79 and 81% of the $Fe^{3+}$ in spinel and garnet lherzolite, respectively, with opx and cpx hosting 48% and 31%, respectively, of the total $Fe^{3+}$ in spinel peridotite. To better understand partitioning of $Fe^{3+}$ between pyroxene and melt we conducted experiments at 100 KPa with $f_{O2}$ controlled by CO-$CO_2$ gas mixes between ΔQFM -1.19 to +2.06 in a system containing andesitic melt saturated with opx or cpx only. To produce large (100-150 μm), homogeneous pyroxenes, we employed a dynamic cooling technique with a 5-10°C/h cooling rate, and initial and final dwell temperatures 5-10°C and 20-30°C super and sub-liquidus, respectively. Resulting pyroxene crystals have absolute variation in $Al_2O_3$ and $TiO_2$ <0.05 wt.% and <0.02 wt.%, respectively. $Fe^{3+}/Fe^T$ in pyroxenes and quenched glass were measured by XANES. We used a newly developed XANES calibration for cpx and opx by only selecting spectra with X-ray vibrating on the optic axial plane at 50±5° to the crystallographic *c* axis. Values of $D_{Fe^{3+}}^{cpx/melt}$ increase from 0.03 to 0.53 as $f_{O2}$ increases from ΔQFM -0.44 to 2.06, while $D_{Fe^{3+}}^{opx/melt}$ remains unchanged at 0.26 between ΔQFM -1.19 to +1.37. In comparison to natural peridotitic pyroxenes, $Fe^{3+}/Fe^T$ in pyroxenes crystallized in this study are lower at similar $f_{O2}$, presumably owing to lower $Al^{3+}$ contents. This study shows that the existing thermodynamic models implemented in pMELTS and Perple_X over-predict the stability of $Fe^{3+}$ in pyroxenes, causing an anomalous reduced character to spinel peridotites at calculated conditions of MORB genesis.

# Declarations

## Funding

The work is supported by the National Science Foundation grants EAR1426772 and EAR2016215

## Conflicts of interest

The authors have no conflicts of interest or competing interests in this study.

## Availability of data

Data will be available upon request to the corresponding author.

## Introduction

Oxygen fugacity ($f_{O2}$) is an intensive thermodynamic variable that influences igneous phase equilibria (Carmichael and Ghiorso 1986), partitioning of multivalent trace elements (Papike et al. 2013), depths of volatile induced melting (Dasgupta and Hirschmann 2010; Rohrbach and Schmidt 2011; Dasgupta 2013; Stagno et al. 2013) and possible saturation of metallic alloy in the mantle (O'Neill and Wall 1987; Rohrbach et al. 2011). In processes associated with partial melting of the mantle and basaltic magmatism, the principal expression of $f_{O2}$ is variation in the ratio of $Fe^{3+}$ and $Fe^{2+}$ in silicate melts and in coexisting mafic minerals (Carmichael and Ghiorso 1990). Consequently, partitioning of $Fe^{3+}$ between mantle minerals and partial melt are crucial parameters necessary to understand relationships between $f_{O2}$, $Fe^{3+}/Fe^{2+}$ ratios in igneous phases, and redox-related processes in basalts and in the upper mantle.

Enrichments in $Fe^{3+}/Fe^{T}$ in Oceanic Island Basalts (OIB) relative to Mid-Ocean Ridge Basalts (MORB) or between MORB in different ridge localities are potential indicators of variable contributions from oxidized recycled sources or of volatiles (Cottrell and Kelley 2011, 2013; Shorttle et al. 2015; Brounce et al. 2017; Moussallam et al. 2019). Variations in $Fe^{3+}/Fe^{T}$ may also be induced by differences in melt fraction because $Fe^{3+}$ is mildly incompatible during partial melting (Holloway 1998; McCanta et al. 2004; Mallmann and O'Neill 2009; Sorbadere et al. 2018). In global compilations of MORB, fractionation-corrected $Fe_2O_3$ correlate negatively with incompatible elements, such as, $Na_2O$ (Bézos and Humler 2005; Cottrell and Kelley 2011) and with isotopic indicators of source enrichment, which favors a connection between source $f_{O2}$ and mantle composition (Cottrell and Kelley 2013). Ocean island basalts from Hawaii, Iceland, the Canaries, and Cape Verde have elevated $Fe^{3+}/Fe^{T}$ compared to MORB (Shorttle et al. 2015; Brounce et al. 2017; Moussallam et al. 2019). This has been attributed to the more oxidized sources of OIB, possibly owing to crustal recycling (Shorttle et al. 2015; Moussallam et al. 2019),but it could also be a product of $Fe^{3+}$ enrichments resulting from melting at higher pressures (Stolper et al. in press) or smaller melt fractions.

MORB and abyssal peridotites are believed to be, respectively, melts and residues of partial melting at mid-ocean ridges (Falloon and Green 1987; Hirose and Kushiro 1993), though each also experience additional processes such as crystal fractionation and melt-rock reaction (Kelemen et al. 1992; Langmuir et al. 1992). MORB record a restricted range of log $f_{O2}$ between ΔQFM (quartz-fayalite-magnetite) (Frost 1991) -0.8 to +0.2 at 100 KPa and 1200° C (Bézos and Humler 2005; Cottrell and Kelley 2011; Berry et al. 2018; Zhang et al. 2018), but abyssal peridotites record a larger amplitude in upper mantle $f_{O2}$ between ΔQFM -1.6 to +2 (Bryndzia and Wood 1990; Birner et al. 2018; Cottrell et al. in press). While a wealth of information is available on $f_{O2}$ and associated $Fe^{3+}/Fe^T$ in basalts and residual peridotites, it is unclear how their redox states relate to one another through partial melting processes.

Although the $Fe^{3+}$ contents of erupted basalts are well constrained by observation, the same is not true for their mantle sources, as different approaches lead to estimates of upper mantle $Fe^{3+}/Fe^T$, that vary by a factor of 3, from ~0.02 to 0.06. Estimates of $Fe^{3+}/Fe^T$ for fertile peridotite derived from observed $Fe_2O_3$ in xenoliths range from 0.02-0.035 (Canil et al. 1994; Canil and O'Neill 1996; Woodland et al. 2006). Calculations of mantle source $Fe^{3+}/Fe^T$ based on inversion of MORB compositions suggest either similar (0.02-0.03; O'Neill et al. (2018)) or greater (>0.035; Cottrell and Kelley (2011)) values, but depend strongly on the assumed value of $DFe^{3+}$ peridotite/basalt. In contrast, Gaetani (2016) found forward modeling of peridotite partial melting using pMELTS (Ghiorso et al. 2002) and an empirical model derived from partition coefficients match the $f_{O2}$ recorded by MORB when the source has $Fe^{3+}/Fe^T$ of 0.06. Similarly, thermodynamic calculations of near-solidus spinel peridotite assemblages with $Fe^{3+}/Fe^T$ = 0.03, using both pMELTS and the models of Jennings and Holland (2015) predict $f_{O2}$s that are substantially more reducing than observed for MORB (Stolper et al. in press), which implies either that the actual source has $Fe^{3+}/Fe^T$ significantly greater than 0.03 or that the thermodynamic models do not capture accurately the behavior of $Fe_2O_3$ in peridotite minerals. The uncertainty in convecting mantle $Fe^{3+}/Fe^T$ (from 0.02 to 0.06) translates to depths of redox carbonate melting beneath mid-oceanic ridges ranging from 85 to 210 km (Stagno et al. 2013) and of nominal Fe-Ni alloy precipitation varying from 220 to 325 km (Rohrbach et al. 2011). These underscore the importance of better understanding of the relationship between peridotite $Fe^{3+}$ in basalt source regions and the $f_{O2}$ of resulting partial melts.

In source regions of basalt, $f_{O2}$ is controlled chiefly by the relative stability of $Fe^{3+}$ between melt and peridotite minerals, including spinel, garnet, clinopyroxene, and orthopyroxene (Frost and McCammon 2008). Although reactions involving spinel and garnet exert primary influence on oxygen fugacity (Mattioli and Wood 1988; Ballhaus et al. 1990; Bryndzia and Wood 1990; Gudmundsson and Wood 1995; Miller et al. 2016), opx and cpx store the greatest share of ferric iron both in spinel and garnet lherzolite facies (79 and 81 wt.%, respectively, Fig. 1). Thus, changes in pyroxene mode and composition control $f_{O2}$ indirectly, by limiting the supply of $Fe^{3+}$ available to form $Fe_3O_4$ (magnetite) in spinel or $Fe_3^{2+}Fe_2^{3+}Si_3O_{12}$ (skiagite) in garnet. This is well illustrated by thermodynamic models (Ghiorso et al. 2002; Jennings and Holland 2015), which predict that reactions involving pyroxene produce changes amounting to ~3 orders of magnitude in $f_{O2}$ across the spinel and garnet lherzolite stability field (Stolper et al. in press). Therefore, understanding the energetics of $Fe^{3+}$ substitution in pyroxene is a key to quantifying how total $Fe_2O_3$ and peridotite mineralogy control upper mantle $f_{O2}$.

The thermochemistry of $Fe^{3+}$ substitution in natural pyroxenes has been examined by Luth and Canil (1993), Sack and Ghiorso (1994a, b, c), and Jennings and Holland (2015). Though these models have proved highly useful for forward modeling of the relationship between iron redox and $f_{O2}$ in subsolidus peridotite and during partial melting (Stolper et al. in press; Jennings and Holland 2015; Gaetani 2016; Sorbadere et al. 2018), they are not based on experiments in which $Fe^{3+}$ in pyroxene was analyzed by an accurate method or in which well-equilibrated pyroxenes and melt have been demonstrated to be in equilibrium with other $Fe^{3+}$-bearing phases.

The most comprehensive and widely used thermodynamic models for $Fe^{3+}$ in pyroxenes are those of Sack and Ghiorso (1994a, b, c), which are embedded in the family of MELTS algorithms (Ghiorso and Sack 1995; Ghiorso et al. 2002; Smith and Asimow 2005). Unfortunately, all the $Fe^{3+}$ determinations in pyroxene used to calibrate these models were calculated stoichiometrically from EPMA analyses. Estimation of $Fe^{3+}$ in pyroxene this way is known to be of low reliability (Dyar et al. 1989; Luth and Canil 1993; Sobolev et al. 1999). Further, because the calibration database was filtered to include only pyroxenes for which the mole fraction of an $Fe^{3+}$-bearing component (esseneite or buffonite) exceeded 5%, most constraints come from zoned cpx grains highly enriched in Na and/or Ti, crystallized from oxidized alkalic basalts (Gee and Sack 1988; Sack and Ghiorso 1994c). These may not be well-equilibrated with coexisting melt or applicable to less extreme pyroxene compositions common in peridotite or low-alkali basalts. Calibration of $Fe^{3+}$ in opx was derived entirely by analogy to cpx and the reasonable assumption that $Fe^{3+}$ prefers for cpx over opx by an arbitrary amount (Sack and Ghiorso 1994a, c).

More recently, Jennings and Holland (2015) developed a thermodynamic model for partial melting of peridotite that includes $Fe^{3+}$ in pyroxene as an $Fe^{3+(M1)}$-$Al^{3+(T)}$ coupled substitution. The model was calibrated to produce $Fe^{3+}$ concentrations in subsolidus peridotite similar to that observed in xenoliths and is also guided by $Fe^{3+}$ partitioning between xenolith garnet and cpx. Though the model is not based on experiments and has less crystal chemical and thermodynamic complexity than that of Sack and Ghiorso (1994a, b, c), it has the advantages that it is calibrated from $Fe^{3+}$ measured by Mössbauer spectroscopy and that it should be tuned to mantle-composition pyroxenes.

Here we report the results of a series of crystallization experiments at 100 kPa that determine $Fe^{3+}$ partition coefficient between opx/melt and cpx/melt under controlled $f_{O2}$ conditions. There are two previous experimental studies on partitioning of $Fe^{3+}$ between pyroxene and melt, including one conducted using Fe-rich Martian basalts (McCanta et al. 2004) and another in a simplified haplobasalt (CaO-$Al_2O_3$-MgO-$SiO_2$ + minor Fe) (Mallmann and O'Neill 2009). Both indicate that $Fe^{3+}$ is incompatible in pyroxene, with the former showing a systematic increase of augite/melt with $f_{O2}$. The latter (Mallmann and O'Neill 2009) inferred the $Fe^{3+}$ content of pyroxene indirectly from variations of the bulk FeO* pyroxene/melt partition coefficient with $f_{O2}$, and from that inferred values of 0.45±0.02 and 0.20±0.02, for cpx/melt and opx/melt, respectively. By design, this method cannot resolve the effects of $f_{O2}$ on $D_{Fe^{3+}}^{pyx/melt}$.

*In situ* quantification of $Fe^{3+}$ in pyroxenes by XANES is complicated by significant spectral anisotropy in the polarized synchrotron beam. Dyar et al. (2002) recommended collection of XANES spectra of anisotropic minerals along optical *y* axes by orienting single crystals on a spindle stage, and this has been developed for analyses of amphiboles (Dyar et al. 2016) and in a preliminary calibration of pyroxenes (Steven et al. 2019). Unfortunately, this approach is impractical for experimental samples in which small (~100 μm) crystals are embedded in glass. Analysis of oriented single crystals also does not facilitate core-rim transects, which are needed to verify homogeneity of experimentally-grown crystals. XANES analysis of anisotropic minerals in thin section can be facilitated by measuring grains with known orientation (e.g., Martin et al. (2017)). To analyze multiple pyroxene grains from each experiment with intact textural relations, we collected XANES spectra of the pyroxenes in suitable crystallographic orientations determined by EBSD.

## Experimental methods

### Starting material

The experiments employed three different starting materials, SM#3, SM#6, and SM#7. These were selected to be saturated in cpx or opx and are based on glass compositions reported by Grove and Juster (1989) with coexisting pyroxene compositions added until the bulk composition became saturated by opx/cpx only. The compositions were prepared from reagent grade oxides ($SiO_2$, $TiO_2$, $Al_2O_3$, $Cr_2O_3$, $Fe_2O_3$, MgO) and carbonates ($CaCO_3$, $Na_2CO_3$, $K_2CO_3$). Oxides other than $Fe_2O_3$ were dried overnight at 1000° C in a box furnace, $Fe_2O_3$ was dried for 1 hour at 800° C, and carbonates were dried overnight at 400° C. The reagents were then weighed and mixed under acetone by grinding with a mortar and pestle until the grinding sound faded. The acetone was then evaporated, and the mixture was decarbonated by heating to 1100°C at 100 °C/hour, followed by a dwell of 24 hours. $Fe_2O_3$ was added afterwards and the powder was mixed by the same methods for 2 hours to ensure homogeneity. The resulting material was then reduced in a horizontal gas mixing furnace at $f_{O2}$ near QFM at 1000° C for 12 hours. To evaluate starting material bulk compositions, aliquots were melted in a vertical gas mixing furnace at 1400° C at a $f_{O2}$ conditions similar to QFM and the resulting glasses were analyzed by were analyzed by electron microprobe (Table 1).

### Experimental procedure

Experiments were conducted in a Deltech VT28 vertical gas mixing furnace at the University of Minnesota. Oxygen fugacity was controlled by flowing a mixture of $CO$-$CO_2$ gas and was measured with a SIRO2 C700 + Solid Zirconia Electrolyte oxygen sensor using air as the reference gas. The sensor EMF was calibrated against the Ni-NiO buffer (O'Neill 1987) by controlled oxidation of a 0.002" diameter Ni wire. The $f_{O2}$ is accurate to within 0.05 log units. Temperature was monitored by a Type S ($Pt_{90}Rh_{10}/Pt_{100}$) thermocouple and temperature uncertainties are ±4° as judged by melting of 0.004" diameter gold wire.

Approximately 20-30 mg of starting mix was mixed with polyvinyl acetate (PVA) and loaded on a 0.005" diameter Pt or Re loop. Growing large homogeneous pyroxenes is challenging owing to the slow diffusion of trivalent cations such as $Al^{3+}$ and $Cr^{3+}$, which promote sector zoning and heterogeneous trivalent cation concentrations (Baker and Grove 1985; Gee and Sack 1988; Hart and Dunn 1993; Hack et al. 1994; Pertermann and Hirschmann 2002; Lofgren et al. 2006). However, we developed a dynamic crystallization technique that produces near-homogeneous pyroxenes (<0.2 wt.% variation in $Al_2O_3$) with typical dimensions of 100-150 μm and well-developed crystal faces. Critical elements of the technique are (1) Pyroxene should be the liquidus phase, as the presence of early olivine (and presumably spinel) offers nucleation centers for the growth of abundant small pyroxenes. This guided our choice of starting mixtures towards more silicic (basaltic andesite to andesite) compositions, rather than more mafic melts which generally do not have pyroxene on the liquidus at 100 kPa. (2) Brief reconnaissance experiments (1-2 hours) were performed to identify the liquidus temperature of each starting composition at $f_{O2}$ intervals of 1 log unit. (3) The charge was heated to 5°-10° above the liquidus and held for 1-2 hours. Longer durations or greater temperatures destroyed crystal nuclei and produced unsatisfactory crystal textures. (4) The charge was then slowly cooled at 5-10° C/hour to ~20-30° below the liquidus and held for 48 hours. Procedures for steps 3 and 4 were established by empirical exploration of the effects of different cooling rates and final dwell times on texture and composition of resultant pyroxene crystals.

Charges were drop quenched in water by fusing the Pt quench wire by resistive heating. Three of the experiments failed to drop quench and were quenched quickly by taking the entire assembly out and submerging in water. Nevertheless, no indication of quench crystals or oxidation rims were observed for these charges. Experimental charges were equilibrated at $f_{O2}$s ranging from ΔQFM -1.19 to +2.06. The CO-$CO_2$ ratio was adjusted during dynamic cooling to maintain a constant ΔQFM throughout the experiment. All experimental conditions are listed in Table 2. Experimental charges were cut into halves using a tungsten wire saw and were polished to 0.25 μm for microanalysis.

## Analytical methods

### Electron microprobe

Electron imaging and major element analyses of the pyroxenes and glass were conducted using a JEOL-JXA 8530FPlus field-emission gun electron microprobe at the University of Minnesota. Quantitative analyses were performed using wavelength dispersive spectroscopy (Table 3). All the phases were analyzed using a 15 kV accelerating voltage and 20 nA beam current. The spot size was 1 μm for pyroxene and 5 μm for glasses, with on-peak count times of 20 s for the unknowns and 10 s for the standards. We analyzed *K*α peaks of the following elements: Si, Ti, Al, Cr, Fe (as FeO), Mg, Ca, Na and K. The primary standards were as follows: $MgCr_2O_4$ (chromite) for Mg and Cr; $FeTiO_3$ (ilmenite) for Fe and Ti; $CaAlSi_3O_8$ (anorthite) for Ca and Si from NMNH collection of standards (Jarosewich 1980); and Taylor $NaAlSi_3O_8$ (albite) for Na;

Taylor $KAlSi_3O_8$ (orthoclase) for K. Kakanui hornblende (Jarosewich 1980) was used as a secondary standard. We used the Armstrong/Love-Scott ZAF procedure for matrix correction using Probe for EPMA© software. X-ray intensities of pyroxenes and glasses were monitored for time dependent intensity (TDI) changes for Na Kα, Si *K*α counts, which were corrected using a self-calibrated correction. Large pyroxene grains and melt pools allowed us to analyze 6-8 points per grain and 8-10 points in glass for each experiment.

## Electron backscattered diffraction

To determine the orientation of the XANES-analyzed pyroxene grains with the x-ray vibration direction, EBSD was performed using a JEOL 6500 SEM at the Characterization Facility at the University of Minnesota. Each selected grain was mapped with a 10 μm step size and 1 s dwell time on each spot. Using the mtex-5.2beta2a toolbox (https://mtex-toolbox.github.io/) implemented in MATLAB, EBSD maps of the grains were rotated about the sample normal to desired angles until the EBSD maps matched with that of the grain in each orientation while the XANES spectra were collected.

## X-ray absorption near edge structure (XANES)

We measured $Fe^{3+}/Fe^T$ of the experimental glasses and pyroxenes using synchrotron-based Fe *K*-edge XANES spectra at the 13-IDE beamline at GSECARS, Advanced Photon Source during 4 visits in July 2017, July 2018, March 2019 and February 2020. The spot size of the X-ray beam was 5 X 5 μm and the incident beam flux varied between ~5 X $10^9$-$10^{11}$ photons/s. APS employs a fixed exit, liquid nitrogen cooled double crystal Si (311) monochromator, as described in Zhang et al. (2016). Spectra were collected in fluorescence mode using a Vortex ME4 silicon-drift diode detector array coupled to a high-speed digital spectrometer system (Quantum Xpress3).

We followed the spectral collection procedure of Zhang et al. (2016) for both glasses and pyroxenes, except that we collected data to higher energies. Spectra were collected between 7020-7350 eV in four regions: 5 eV steps between 7020-7105 eV with 1 s dwell; 0.1 eV steps between 7106-7118 eV with 2 s dwell, 1.0 eV steps between 7118-7120 eV with 1 s dwell and 2.0 eV steps from 7120-7350 eV with 1 s dwell. The LW_0 reference glass standard (Cottrell et al. 2009) was analyzed every 6-8 hours to monitor energy drift. The centroid energy of LW_0 was measured to be 7111.96±0.02 ($n$=18) within uncertainty as reported by Zhang et al. (2016). As we observed no drift in LW_0 centroid energy, no time-dependent correction was applied. For each experiment, 2-3 pyroxene grains with 4-6 spots on each crystal were analyzed. 6-8 spectra of glasses were collected for each experiment.

**Measurement of $Fe^{3+}/Fe^{T}$ in glasses**

XANES spectra of the glasses were processed with the Larch XAFS analyses package (Newville 2013). The spectra were corrected for deadtime and for self-absorption using the FLUO algorithm (Haskel 1999) embedded in Larch. The spectra were then edge-step normalized such that the average intensity for the EXAFS region at 7200-7350 eV was assigned a value of 1. We then followed the pre-edge extraction method developed by Cottrell et al. (2009). The pre-edge baseline is fitted using a combination of linear and damped harmonic oscillator functions (DHO) between 7110-7118 eV. The extracted pre-edge is then fitted with two Gaussian sub peaks of ($Fe^{2+}$ and $Fe^{3+}$). Because of the difference in monochromator energy calibration between NSLS and APS of 1.3 eV (Zhang et al. 2016), all glass spectra collected in this study were corrected to a LW_0 centroid energy = 7112.3 eV (Cottrell et al. 2009). Using the intensity and center position of the individual subpeaks, we calculated the area-weighted pre-edge centroid energy and determined $Fe^{3+}/Fe^{T}$ of the glasses from the andesite-derived calibration curve, given in Eqn. 4 of Zhang et al. (2018),

$$\frac{Fe^{3+}}{Fe^{T}} = a_1 + a_2(E-\mu) + a_3(E-\mu)^2 \tag{1}$$

where $E$ is the area weighted pre-edge centroid energy, $\mu = 7113.25$ eV, $a_1 = 0.64116$, $a_2 = 0.77511$, and $a_3 = 0.26251$. The compounded precision uncertainties ($\sigma_{y2}$) are calculated by considering the instrumental uncertainties in measuring the pre-edge centroid energies ($\sigma_x$), uncertainties in coefficients to the quadratic equation and their covariances following the methodology of Zhang et al. (2018),

$$\sigma_{y2} = \sqrt{X^{T} \operatorname{cov} X + {\sigma_{y1}}^{2}} \tag{2}$$

Here, $\sigma_{y1}$ is the uncertainty arising from the instrumental uncertainty in the centroid determination given by, $\sigma_{y1} = \sqrt{(a_2 + 2a_3 x)^2 \sigma_x}$, and $X^{T}$ is a vector given by [1 $x$ $x^2$], and $\operatorname{cov} X$ is a variance-covariance matrix of the coefficients of the quadratic equation given by (Bevington and Robinson, 2003),

$$\operatorname{cov} X = \begin{pmatrix} \sigma^2_{a_1} & \sigma^2_{a_1 a_2} & \sigma^2_{a_1 a_3} \\ \sigma^2_{a_2 a_1} & \sigma^2_{a_2} & \sigma^2_{a_2 a_3} \\ \sigma^2_{a_3 a_1} & \sigma^2_{a_3 a_2} & \sigma^2_{a_3} \end{pmatrix} . \tag{3}$$

**Measurement of $Fe^{3+}/Fe^{T}$ in pyroxene**

High precision measurement of $Fe^{3+}/Fe^{T}$ in pyroxenes by XANES requires development of a standard-based calibration, with the $Fe^{3+}/Fe^{T}$ analyzed by an independent method. *In situ* micro-beam measurement of $Fe^{3+}/Fe^{T}$ in pyroxenes is challenging owing to the directional anisotropy of the X-ray absorption spectra. As shown in the previous studies, orientation induced anisotropy of XANES spectra in cpx and opx contributes ~15-20% (absolute) uncertainty in measured $Fe^{3+}/Fe^{T}$ (Dyar et al. 2002; McCanta et al. 2004).

To improve XANES analyses of $Fe^{3+}/Fe^{T}$ in pyroxene, we developed a Mössbauer-based calibration that accounts for anisotropy using EBSD measurements of pyroxene crystallographic orientation. For calibration, we accumulated a collection of 13 cpx and 3 opx that we established by EPMA to be homogeneous and inclusion free (Supplementary Table 1). To minimize the directional anisotropy of the XANES spectra, we determined the orientation of the pyroxene standards using EBSD. The spectra where the pyroxene crystals are orientated such that the X-ray vibration direction lies in the optic axial plane at an angle of 50±5° to the *c*-axis were selected for calibration. Unlike glasses, pyroxene XAFS spectra show considerable variation in structure up to ~7250 eV; therefore, we normalized the spectra at higher energies (7300-7350 eV) to minimize the influence of normalization in spectral shape. The pre-edge region of the XANES spectra were fitted between 7105-7118 eV. The rising edge was modelled by a combination of linear and Lorentzian functions. After baseline subtraction, the pre-edge was fitted with 3-4 pseudo-Voigt sub-peaks. The requirement of multiple sub-peaks for fitting the pre-edge region is in line with the theoretical considerations of XANES pre-edge modelling of $Fe^{2+}$ and $Fe^{3+}$ in octahedral sites (Westre et al. 1997; De Groot et al. 2009). The set of pyroxenes used for calibration, their analyses, and XANES data collection and fitting methods are provided in a separate study (Rudra and Hirschmann, in preparation).

We fitted the variation of $Fe^{3+}/Fe^{T}$ with area-weighted centroid energy for 13 cpx standards with a weighted least-square linear fit, taking into account the uncertainties in pre-edge centroid energy and Mössbauer determined $Fe^{3+}/Fe^{T}$ (Fig. 2). This has a $r^2$ of 0.985, mean squared weighted deviation (MSWD) of 3.95 and the linear function is given by, $E = m_1 Fe^{3+}/Fe^T + c_1$, where $E$ is the drift-corrected pre-edge centroid energy minus 7112.0 eV, $m_1 = 1.729 \pm 0.168$, $c_1 = 0.205 \pm 0.052$, and covariance = 0.008. The resulting calibration reproduces MÖssbauer determined $Fe^{3+}/Fe^T$ by an average deviation of ±3.6% (absolute), which is comparable to the accuracy of XANES measurements reported for pyroxene by McCanta et al. (2004) (±1-5%) and Martin et al. (2017) (3.2-4.2%) and for amphibole (5.5-6.2%) (Dyar et al. 2016).

Similar to cpx, the variation of pre-edge centroid energy as a function of $Fe^{3+}/Fe^T$ in the 3 opx samples was fitted with a weighted least-square linear fit (Fig. 2). We note that we have few opx standards and that their distribution is strongly biased to low $Fe^{3+}/Fe^T$ ratios. Therefore, nuances in the variation of centroid energy with $Fe^{3+}/Fe^T$ for opx are probably not captured in this calibration, and in the future, additional samples with higher ferric iron content will be required. At best, this preliminary opx calibration is only applicable for $Fe^{3+}/Fe^T$ between 0.02-0.12, has an $r^2$ of 0.971, MSWD = 0.27 and is given by, $E = m_2 Fe^{3+}/Fe^T + c_2$, where $E$ is the drift-corrected centroid energy minus 7112.0 eV. $m_2 = 5.671 \pm 1.003$, $c_2 = -0.115 \pm 0.104$, covariance = 0.098.

To measure the $Fe^{3+}/Fe^T$ in randomly oriented unknown pyroxenes, multiple grains were analyzed for each experiment. Using an automated rotating stage or by manual rotation, spectra for each grain were collected in three different orientations differing by 45° (Fig. 3a). Following beamline observation, the crystallographic orientations of each pyroxene crystal relative to this X-ray geometry were measured by EBSD. Spectra for which the incident X-ray was vibrating in the *a*-*c* plane (optic axial plane) of the pyroxenes at an angle of 50±5° to the *c* axis were chosen. The selected spectra were then pre-processed by deadtime correction, edge-step normalization, and self-absorption correction implemented in Larch. The pre-edge centroid energy of each spectrum was determined by fitting the baseline with a combination of linear and Lorentzian functions and fitting 3-4 pseudo-Voigt subpeaks between 7105-7118 eV (Fig. 3b). The uncertainties in measured $Fe^{3+}/Fe^{T}$ ($\sigma_{y}$) for the unknown pyroxenes are calculated by incorporating the instrumental uncertainties in measured centroid position and the uncertainties of the coefficients and their covariance, given by the equation,

$$\sigma_{y} = \frac{1}{m}\sqrt{\sigma_{E}^{2} + \sigma_{c}^{2} + \frac{\sigma_{m}^{2}(c-E)^{2}}{m^{2}} + \frac{2\sigma_{mc}^{2}(E-c)}{m}} \tag{4}$$

Where $\sigma_{E}$ is the instrumental uncertainty in XANES pre-edge centroid measurement, $\sigma_{c}$, $\sigma_{m}$, are the variances of the slope and intercept, respectively, of the linear equation and $\sigma_{mc}$ is the covariance between them.

## Results

### Textural properties of pyroxene and melt

The experiments produced large (80-150 μm), euhedral to subhedral, near-homogeneous pyroxene crystals across the entire $f_{O2}$ range (Fig. 4). Starting materials, SM#3 and SM#7 only produced cpx+glass assemblages; SM#6 produced opx+glass. All experiments had large areas of quenched glass, free from micro-crystals, allowing microbeam (XANES, EPMA) analysis of many individual spots. Our preferred formula of dynamic cooling produced pyroxenes with small variations in $Al_2O_3$ and $TiO_2$ concentrations across the crystals as can be seen in the BSE image and quantitative X-ray maps (Fig. 5a, b, d). Most of the pyroxene crystals have distinct cores characterized by small enhancement of $Al_2O_3$ (+0.3 wt. %) and $TiO_2$ (+0.06 wt. %) concentrations, which served as early nucleation sites for the rims to grow during cooling. In contrast, the crystals have more homogenized distribution of FeO* with no distinct core to rim zoning (Fig. 5c). The crystals developed thick and near-homogeneous rims with absolute concentrations of $Al_2O_3$ and $TiO_2$ varying by <0.05 wt.% and <0.02 wt.% respectively across a length scale of 40-50 μm. These are taken as the compositions of pyroxene that equilibrated with adjacent melt and are sufficiently large to facilitate collection of many individual XANES and EPMA analyses (Fig. 5e).

The nearly homogeneous crystals grown by our preferred method can be compared with commonly employed techniques of growing crystals in pyroxene/melt elemental partitioning experiments (Colson et al. 1988; Beattie et al. 1991; Gaetani and Grove 1995; Blundy and Dalton 2000; Pertermann and Hirschmann 2002; Gaetani et al. 2003). Isothermal experiments and dynamic procedures with any combination of the following steps - faster cooling rates (45°-60° C/h), smaller final dwell times (12-24 h), and/or with initial dwell temperatures appreciably below (40-50° C) the liquidus temperature produced abundant (mode 60-70%), small (5-20 μm) crystals showing concentric growth zoning in multiple major elements. We surmise that the large difference between the initial dwell temperature compared to the liquidus preserves abundant nucleation sites for the crystals to grow later during programmed cooling, which lead to numerous, small crystals. In addition, relatively short final dwell times did not allow crystals to homogenize, which preserved concentric growth zoning. For example, an experiment, VF116, with faster cooling rate (40° C/h), shorter final dwell time (24 h) and initial dwell temperature of 45° C below the liquidus produced small, anhedral pyroxene crystals with prominent core to rim zoning of multiple major elements (Fig. 6a, b). The transect across a cpx crystal from this experiment shows associated sinusoidal variation of FeO* and $Al_2O_3$, possibly indicating $Al^{3+}$ controlled substitution of $Fe^{3+}$ in pyroxene structure (Fig. 6c). Large variations in all the oxides, even at scales as small as 2.5 μm, would be particularly problematic during XANES spectra collection, as the typical beam diameter is larger than this. These factors underline the importance of our preferred crystallization formula in obtaining equilibrium elemental partition coefficients for pyroxene/melt partitioning experiments.

**Electron microprobe analysis**

Compositions of opx and cpx vary to different degrees across the $f_{O2}$ range investigated in this study. While the Mg# ($X_{MgO}/(X_{MgO}+X_{FeO^*})$) of cpx vary from 91-92, opx shows a larger range of variation from 80-87, which progressively increases with increasing $f_{O2}$. Cpx shows modest variation in FeO* and $TiO_2$ wt.% between 3.13-3.57 wt.% and 0.11-0.14 wt.% respectively across the 2.5 log units of $f_{O2}$. In contrast, cpx shows larger variation in $Al_2O_3$, $Na_2O$ and CaO between 0.56-1.31 wt.%, 0.17-0.32 wt.% and 19.37-21.61 wt.% respectively, progressively becoming more aluminous and alkali rich with increasing $f_{O2}$. The opx show large variation in FeO* and CaO from 10.02–12.03 wt.% and 0.41-1.45 wt.% respectively, progressively becoming enriched in total Fe and Ca with increasing furnace $f_{O2}$. The opx also show modest variation in $Al_2O_3$ wt.% and $TiO_2$ between 0.33-0.43 wt.% and 0.14-0.17 wt.%., respectively, suggesting the paucity of charge coupled substitutions in stabilizing tri and tetra-valent cations in opx.

Both the opx and cpx saturated glasses show small variation in several oxides across the $f_{O2}$ range considered. The analyzed FeO* in the glasses have high precision (usually ±1-1.5% relative) and are believed to have high accuracy as the FeO* of the secondary standard was reproduced with an average discrepancy of ±0.07 wt.% with a largest discrepancy of 0.1 wt.%. In the cpx and opx saturated glasses FeO* varies between 6.46-7.37 wt.% and 9.16-9.77 wt.% respectively. In addition to FeO*, other oxides, such as, $SiO_2$, $Al_2O_3$, $TiO_2$ show modest variation in $f_{O2}$ across the range investigated.

**XANES analysis**

In the cpx-saturated experiments, $Fe^{3+}/Fe^T$ in the experimental glasses increases from between 0.151 ± 0.016 to 0.309 ± 0.012 as the $f_{O2}$ increases from ΔQFM -0.44 to 2.06. In the opx-saturated experiments, the $Fe^{3+}/Fe^T$ changes from 0.118 ± 0.019 to 0.229 ± 0.013 with increasing $f_{O2}$ from ΔQFM -1.19 to 1.37. Measured centroid energies and the calculated $Fe^{3+}/Fe^T$ for all the experimental glasses are given in Table 4. The measured $Fe^{3+}/Fe^T$ in the glasses agrees well with the predicted value by Kress and Carmichael (1991) (Fig. 7). Owing to compositional differences, originating through the composition dependent terms in the Eqn. 7 of Kress and Carmichael (1991), the model predicts that the opx-saturated glasses have systematically lower $Fe^{3+}/Fe^T$ at a similar $f_{O2}$ than the cpx-saturated glasses. XANES analyses of the glasses also conform to the theoretical model and show that the measured $Fe^{3+}/Fe^T$ in opx saturated glasses are lower than their cpx saturated counterpart by 2-6% (absolute) in experiments equilibrated at similar $f_{O2}$.

The newly developed XANES based calibration for cpx and opx enable us to measure $Fe^{3+}/Fe^T$ in the experimental pyroxene crystals. $Fe^{3+}/Fe^T$ in cpx increases from 0.02±0.03 to 0.30±0.05 as the $f_{O2}$ increases from ΔQFM -0.44 to 2.06. Within a single crystal, $Fe^{3+}/ Fe^T$ varies by 3-5% (relative), and the variation is within analytical uncertainty of individual analysis. Inter-crystalline variation in $Fe^{3+}/Fe^T$ from a single experiment is 6-9% (relative), which could be a result of small differences in major element concentrations and differences in the orientation of the crystals with respect to X-ray vibration direction within the permissible limit of our selection criteria. $Fe^{3+}/ Fe^T$ in the opx changes from 0.03±0.02 to 0.06±0.02 as the $f_{O2}$ increases from ΔQFM -1.19 to 1.37. Within an experiment, inter-crystalline and intra-crystalline variations in opx $Fe^{3+}/ Fe^T$ are 8-11% and 9-11% respectively. Values of the pre-edge centroid energies and corresponding $Fe^{3+}/Fe^T$ are given in Table 4.

Pyroxene/melt partition coefficients can be calculated from the $Fe^{3+}$ concentrations in pyroxene and coexisting melt. Uncertainties in the calculated $D_{Fe^{3+}}^{pyx/melt}$ derive from the uncertainties in $Fe_2O_3$ wt.% both in the glass and pyroxene following equation

$$\sigma_{D_{pyx/Gl}^{Fe3+}} = \left(\frac{1}{Fe_2O_3^{\ Gl}}\right)\sqrt{\sigma^2_{Fe_2O_3^{pyx}} + \sigma^2_{Fe_2O_3^{Gl}}\left(D_{pyx/Gl}^{Fe3+}\right)^2} \quad (5)$$

where and $\sigma_{Fe_2O_3}$ are uncertainties in the $Fe_2O_3$ wt.% in glasses and pyroxenes calculated from the uncertainties in the $Fe^{3+}/Fe^T$ and FeO* of the respective phases using the equation,

$$\sigma_{Fe_2O_3} = \left(\frac{1}{0.8998}\right)\sqrt{\sigma^2_{FeO^*}\left(\frac{Fe^{3+}}{Fe^T}\right)^2 + \sigma^2_{\left(\frac{Fe^{3+}}{Fe^T}\right)}\left(FeO^*\right)^2} \quad (6)$$

Our results show that the $D_{Fe^{3+}}^{cpx/melt}$ increases continuously from 0.03-0.53 with increasing $f_{O2}$, suggesting that $Fe^{3+}$ becomes progressively more stable in cpx as $Fe^{3+}$ becomes more abundant in the system. In contrast, $D_{Fe^{3+}}^{opx/melt}$ remains unchanged within analytical uncertainly at an average value of 0.26 across the entire range of $f_{O2}$ (Fig. 8). Values of partition coefficients for all the experiments are given in Table 4.

# Discussions

## Comparison to previous determinations

In an experimental study of basalts related to Martian (SNC) meteorites, McCanta et al. (2004) also observed increases in $D_{Fe^{3+}}^{augite/melt}$ with increasing $f_{O2}$, although the absolute values of their partition coefficients are larger by a factor of 2 than those from the present study (Fig. 9). In addition, McCanta et al. (2004) found values of $D_{Fe^{3+}}^{pigeonite/melt}$ that are higher than $D_{Fe^{3+}}^{augite/melt}$ from either study and these correlate negatively with $f_{O2}$. The observation of preferential concentrations of $Fe^{3+}$ in pigeonite compared to augite at comparable $f_{O2}$ is unexpected on the basis of crystal chemical considerations, as noted by McCanta et al. (2004). Differences between the results of the two studies might arise from different bulk compositions employed (18-20 vs. 7-10 wt.% FeO*, 48-50 vs. 56-61 wt.% $SiO_2$). They may also arise from differences in experimental approaches. McCanta et al. (2004) grew pyroxenes from melt with dynamic cooling experiments that employed a rapid cooing rate (60° C/h). As shown in Fig. 6, such high cooling rates are not conducive to well-equilibrated pyroxene textures. In particular, rapid cooling is known to affect concentrations of tri and tetra-valent cations in pyroxene (Gamble and Taylor, 1980).

Compared to the partition coefficients determined by Mallmann and O'Neill (2009), the values determined in this study at ΔQFM > 0.5 agree well, although their experiments were not designed to determine the $f_{O2}$ dependence of partitioning. However, we note that values of $D^{pyx/melt}_{Fe^{3+}}$ in the Mallmann and O'Neill (2009) study derive from estimates of $Fe^{3+}$ in pyroxenes deduced from stoichiometry of EPMA analyses, and therefore may have large uncertainties.

**Comparison of pyroxene $Fe^{3+}/Fe^T$ between experiments and xenoliths**

To our knowledge, there are no systematic studies of $Fe^{3+}/Fe^T$ in natural volcanic pyroxene as a function of $f_{O2}$. However, the $Fe^{3+}/Fe^T$ of the experimental pyroxenes can be compared to those in natural peridotitic pyroxenes for which $f_{O2}$ is also known by spinel oxybarometry (Dyar et al. 1989; Luth and Canil 1993; Canil and O'Neill 1996; Woodland et al. 2006), which means that they have significant compositional differences with those crystallized from andesite at 100 kPa. For the peridotites, we recalculate oxygen fugacities from ol-opx-sp following the methodology of Mattioli and Wood (1988) and Wood and Virgo (1989) and including a pressure correction term determined from the standard state volume change of the reaction (Davis et al.

2017). The relative $f_{O2}$ is calculated with respect to QFM from the parameterization of Frost (1991). Site occupancies of Fe in opx ($X_{M1}^{Fe}, X_{M2}^{Fe}$) are calculated following the methodology of Wood and Banno (1973); and activity of magnetite in spinel ($a_{sp}^{Fe_3O_4}$) following the Nell and Wood model (Ionov and Wood 1992). We estimated the temperature of equilibrium based on the Fe-Mg exchange reaction between olivine and spinel (Li et al. 1995), as these phases are also involved in the oxybarometric reaction (Birner et al. 2018). Owing to the lack of a suitable geobarometer, we assumed a mean pressure of 1.5 GPa for the spinel-peridotite xenoliths. As shown in Fig. 10, ratios of $Fe^{3+}/Fe^{T}$ in cpx and opx are comparable in the experiments and in natural peridotites, though at similar $f_{O2}$ the former may have lower ratios. The differences could be owing to multiple factors, including temperature, pressure and pyroxene composition. In particular, the comparatively low $Al_2O_3$ content of the experimental pyroxenes (0.56-1.31 wt.%) could be an important factor. However, systematic understanding of the roles of intensive and compositional variables on $Fe^{3+}$ substitution in pyroxene is beyond the scope of this study.

## Comparison of $D\mathrm{Fe}^{3+}$ between experiments and thermodynamic models

The experimental results can also be compared to predictions from thermodynamic models. Here we examine $D_{Fe^{3+}}^{pyx/melt}$ in two widely used parameterizations for $Fe^{3+}$ in pyroxenes by Sack and Ghiorso (1994a, d, c), and Jennings and Holland (2015), implemented in pMELTS (Ghiorso et al. 2002) and Perple_X (Connolly, 1990), respectively (Fig. 11). The calculations are done at 100 kPa and 1230° C and 1208° C, under oxygen buffered conditions between ΔQFM -2 to +2, for cpx/melt and opx/melt pairs, respectively. For pMELTS, the bulk compositions used for opx and cpx saturation are the same as the compositions used in this study. For calculations in Perple_X, we used the basaltic LOOS composition used in Davis and Cottrell (2018), as the melt model of Jennings and Holland (2015) is not calibrated for andesitic liquids. Both models predict values of $D_{Fe^{3+}}^{pyx/melt}$ that are greater than those evident from the experiments in this study.

Values of $D_{Fe^{3+}}^{pyx/melt}$ calculated with pMELTS increase with $f_{O2}$, as also observed from experiments, but absolute values are greater - from 0.34-0.83 for $D_{Fe^{3+}}^{cpx/melt}$ and 0.24 to 0.57 for $D_{Fe^{3+}}^{opx/melt}$. Higher values of $D_{Fe^{3+}}^{pyx/melt}$ calculated for the same compositions as the experiments suggest that $Fe^{3+}$ is too stable in pyroxene in the pMELTS models. This might account for the observation that $f_{O2}$ along the solidus of a spinel lherzolite with bulk $Fe^{3+}/Fe^T$ = 0.03 is too reduced to be in equilibrium with MORB (Stolper et al. in press) or equivalently, that matching the $f_{O2}$ of MORB in spinel peridotite requires greater values of $Fe^{3+}/Fe^T$ (0.06; Gaetani 2016).

Values of $D_{Fe^{3+}}^{pyx/melt}$ predicted by the model of Jennings and Holland (2015) are distinct from those from the experiments in three different ways: (1) Both $D_{Fe^{3+}}^{cpx/melt}$ and $D_{Fe^{3+}}^{opx/melt}$ are systematically higher than the experimentally derived partition coefficients. (2) $Fe^{3+}$ behaves as a compatible element in orthopyroxenes, leading to values of $D_{Fe^{3+}}^{opx/melt}$ that are a factor of ~3 higher than $D_{Fe^{3+}}^{cpx/melt}$. This is surprising, as the model is calibrated in part from natural pyroxenes from garnet lherzolites and the $Fe_2O_3$ wt.% in opx in xenoliths are systematically lower compared to cpx at all *P-T* conditions (Dyar et al. 1989; Luth and Canil 1993; Canil and O'Neill 1996; Woodland et al. 2006; Woodland 2009). (3) Calculated values of $D_{Fe^{3+}}^{cpx/melt}$ decrease with increasing $f_{O2}$; although, the dependence is stronger for opx than that of cpx. Although, we used a different liquid composition compared to our starting mixes, the negligible dependence of $D_{Fe^{3+}}^{pyx/melt}$ on melt composition does not change these observations. If applied to spinel peridotites, the large stabilization of $Fe^{3+}$ in pyroxenes predicted by Jennings and Holland (2015) would be expected to dilute the magnetite component in spinel, resulting in lower $f_{O2}$ along the solidus of spinel lherzolite compared that predicted by pMELTS, as observed by Stolper et al. (in press).

## Conclusions

- Opx and cpx are the chief reservoirs of $Fe^{3+}$ in spinel peridotite, hosting approximately 48% and 31%, respectively, of the bulk rock $Fe_2O_3$ wt.%. Therefore, understanding the energetics of $Fe^{3+}$ substitution in pyroxene is crucial to understanding relationships between bulk composition, oxygen fugacity during partial melting of peridotite and for subsolidus conditions.
- Growing large, near-homogeneous pyroxene crystals is challenging, but essential to experimental determination of equilibrium $D_{Fe^{3+}}^{pyx/melt}$. Dynamic cooling experiments with initial dwell temperatures 5-10° above the liquidus followed by slow cooling at 5-10°C/h to a dwell 25-30° below the liquidus for 48 hours produces large (100-150 μm), near homogeneous pyroxenes in equilibrium with the melt.
- *In situ* XANES measurements of pyroxene facilitated by EBSD for orientation determination significantly reduces errors in measured $Fe^{3+}/Fe^T$. By selecting XANES spectra with X-ray vibrating only on the optic axial plane at an angle of 50±5° to the *c* axis reproduces Mössbauer-measured $Fe^{3+}/Fe^T$ at an average deviation of ±3.6% (absolute). This accuracy is comparable to that found in previous studies for which XANES measurements were conducted on optically oriented single crystals (Dyar et al. 2002, McCanta et al. 2004, Dyar et al. 2016).
- $D_{Fe^{3+}}^{cpx/melt}$ continuously increases from 0.03-0.53 as a function of $f_{O2}$, while $D_{Fe^{3+}}^{opx/melt}$ remains unchanged at 0.26 within analytical uncertainty across the $f_{O2}$ range investigated.

- Experimental $D_{Fe^{3+}}^{pyx/melt}$ values in this study agree well partition coefficients determined by Mallmann and O'Neill (2009), but are smaller by a factor of 2 compared to those determined by McCanta et al. (2004). When comparing with natural pyroxenes from peridotite xenoliths and orogenic lherozolite massifs, $Fe^{3+}/Fe^{T}$ in experimentally-grown pyroxenes occur at the lower boundary of the field defined by natural samples. Lastly, comparison to experimentally determined $D_{Fe^{3+}}^{pyx/melt}$ with the thermodynamic models of Ghiorso et al. (1994a, b, c) and Jennings and Holland (2015) reveal that both models over-predict the stability of $Fe^{3+}$ in opx and cpx. This observation potentially resolves the discrepancy of $f_{O2}$ in spinel peridotite between modelled and expected - based on equilibrium with MORB - without requiring unusually large $Fe^{3+}/Fe^{T}$ ratios (Gaetani 2016).

**Figure captions**

**Fig. 1** Average distribution of $Fe^{3+}$ in peridotite for spinel (*n*=52) and garnet (*n*=75) peridotite (Dyar et al. 1989; 1992; McGuire et al. 1991; Canil et al. 1994; Canil & O'Neill 1996; Woodland & Koch 2003; Woodland et al. 2006; Woodland 2009; Nimis et al. 2015) derived from modes and $Fe^{3+}$ in coexisting peridotite minerals from xenoliths and orogenic peridotites. Though the distribution of $Fe^{3+}$ between phases in peridotite varies with temperature and pressure (Luth & Canil 1993; Canil & O'Neill 1996; Gaetani 2016; Birner et al. 2018; Davis & Cottrell 2018), pyroxene will be the dominant host of $Fe^{3+}$, at least at depths <250 km. As the asthenosphere is hotter than typical xenoliths, the fraction of $Fe^{3+}$ in spinel and garnet in the convecting mantle should be respectively less (Gaetani 2016; Stolper et al. in press) and greater (Luth & Canil 1993; Nimis et al. 2015) than depicted here

**Fig. 2** Mössbauer based XANES calibration of $Fe^{3+}/Fe^{T}$ for cpx and opx. Mössbauer-determined $Fe^{3+}/Fe^{2+}$ ratios are corrected for unequal recoilless fractions of $Fe^{3+}$ and $Fe^{2+}$ at room temperature, *c*, of 1.2 (De Grave and Van Alboom 1991, Eeckhout and De Grave 2003). Both for cpx and opx, only the spectra where the X-ray is vibrating in the appropriate direction with respect to the crystallographic axes are considered for calibration (see text). The red dashed line is the linear fit to the cpx data points with reduced $\chi^2 = 3.95$. The calibration for opx is limited to $Fe^{3+}/Fe^{T}$ between 0.02-0.12. The green dashed line is a linear fit to the opx data points. All error bars are $1\sigma$ standard deviation

**Fig. 3** (a) Reflected light image of basal section of a cpx obtained during XANES data collection at the synchrotron beamline. The yellow circle marks the position of the X-ray beam. The beam is strongly polarized in the horizontal plane denoted by the parallelogram, which is at an angle of 90° to the plane of the image. The dashed lines mark the incoming and outgoing X-ray directions on the horizontal plane. (b) Pre-edge region of XANES spectrum of a clinopyroxene in VF118. The blue curve is the pre-edge region of the normalized XANES spectrum, and the grey dashed lines represent the baseline and the pseudo-Voigt sub-peak fits

**Fig. 4** BSE micrograph of VF117 (1228°C, QFM+0.86) showing euhedral-subhedral cpx coexisting with melt. Note that cpx in VF117 have melt inclusions, but these are avoidable during analysis. The experimental charge has large melt pools free from microcrystals allowing multiple uncontaminated micro-beam analyses

**Fig. 5** Micrograph of cpx embedded in andesitic glass from experiment VF122, demonstrating textural features of a well-equilibrated pyroxene grain grown from melt by the favored method (cooling rate: 5° C/h, dwell time: 48 h, first dwell temperature ~5° C below the liquidus) embedded in andesitic glass. (a) BSE electron image (30nA, 15kV) showing euhedral crystal

habit with negligible major element heterogeneity. (b, c, d) Quantitative WDS X-ray maps (80nA, 15kV) of $TiO_2$, FeO* and $Al_2O_3$ of the same crystal. $TiO_2$ and $Al_2O_3$ maps show a distinct core with modestly elevated concentrations of both the oxides compared to the rim (see text for more details). Compared to the other oxides, FeO* is more homogeneously distributed across the crystal with no distinct core to rim zoning. The surrounding glass has large areas free of microcrystals as evident from the homogeneous distribution of oxides. We use a logarithmic color scale in the maps to highlight the variation in concentrations of oxides within the crystal. The color bars on the right show concentrations of oxides. (e) variation of all the three major elements across the crystal along the indicated X-Y transect. As can be seen in the maps, $Al_2O_3$ and $TiO_2$ concentrations are elevated in the core compared to the rim with absolute differences of 0.4 wt.% and 0.1 wt.% respectively. The concentration profiles become flatter in the rims across a length scale of 40-50 μm. FeO* doesn't show a systematic difference between core and rim and varies by ~0.05 wt.%

**Fig. 6** Micrograph of cpx crystals in andesitic glass from experiment VF116, showing heterogeneous textures resulting from more rapid cooling rates (cooling rate: 40° C/h, dwell time: 24 h, initial dwell temperature ~45° C below the liquidus) (a, b) FeO* and $Al_2O_3$ maps (80nA, 15 kV) of aggregate of cpx crystals from VF116. Crystals are subhedral to anhedral and range between 10-20 μm. Crystals are marked by prominent core to rim zoning in major elements (see text for more details). The logarithmic color scales on the right show absolute concentration of the oxides. (c) Variation of $Al_2O_3$, $TiO_2$ and FeO* across one of the cpx crystals along the indicated X-Y transect. $Al_2O_3$ and FeO* show concomitant variations in a sinusoidal pattern possibly indicating that $Fe^{3+}$ substitution is controlled by $Al^{3+}$. Absolute concentrations of FeO* and $Al_2O_3$ vary by ~1.9 wt.% and 0.6 wt.% across the length scale of ~3 μm. $TiO_2$ wt.%

varies by 0.7 wt.% across the crystal and the pattern of variation is antithetic to variation of other oxides

**Fig. 7** Variation of $Fe^{3+}/Fe^{T}$ in glasses as a function of ΔQFM measured by XANES spectroscopy. The dashed lines show upper and lower ranges of $Fe^{3+}/Fe^{T}$ at 1230° C predicted by Kress and Carmichael (1991) for the range of glass compositions in the experiments. Some of the measured data points fall outside the range of predicted $Fe^{3+}/Fe^{T}$ owing to the difference in final dwell temperature between those experiments and that of the model calculation. All uncertainties are 1σ standard deviations

**Fig. 8** Experimentally determined $Fe^{3+}$ partition coefficients between opx/melt and cpx/melt as a function of relative $f_{O2}$. The average $D_{Fe^{3+}}^{opx/melt}$ is 0.26 and is constant within analytical uncertainty across the range of $f_{O2}$ investigated here. $D_{Fe^{3+}}^{cpx/melt}$ shows a positive correlation with ΔQFM suggesting stabilization of $Fe^{3+}$ in cpx as $Fe^{3+}$ becomes more abundant in the system. All uncertainties are 1σ standard deviation

**Fig. 9** Comparison of $D_{Fe^{3+}}^{pyx/melt}$ between this work and two previous studies. $D_{Fe^{3+}}^{augite/melt}$ in Martian basalts from McCanta et al. (2004) are larger by a factor of ~2 than those from the present work but both studies show a positive dependence on $f_{O2}$. The grey horizontal bars are $D_{Fe^{3+}}^{pyx/melt}$ values from Mallmann and O'Neill (2009) in a CFMAS system. They match well with the $D_{Fe^{3+}}^{pyx/melt}$ values in this study at ΔQFM > 0.5

**Fig. 10** Comparison of $Fe^{3+}/Fe^{T}$ in cpx and opx between experiments in this study and spinel lherzolites from continental lithospheric xenoliths and orogenic massifs. (Woodland et al. 2006; Dyar et al. 1989; Canil and O'Neill 1996; Luth and Canil 1993). $f_{O2}$ is recalculated using ol-opx-

sp by using the ol-sp equilibration temperature and an assumed fixed pressure of 1.5 GPa (see text for more details). At similar $f_{O2}$, $Fe^{3+}/Fe^{T}$ is more enriched in opx and cpx from xenoliths compared to that of experiments. This difference in $Fe^{3+}/Fe^{T}$ could reflect an effect of increasing pressure and decreasing temperature in preferential partitioning of $Fe^{3+}$ in pyroxene compared to spinel in the more aluminous peridotitic pyroxenes

**Fig. 11** Comparison of $D_{Fe^{3+}}^{pyx/melt}$ between experimental results and the thermodynamic models of Ghiorso et al. (1994a, b, c) and Jennings and Holland (2015). The two models are implemented in pMELTS (Ghiorso et al. 2002) and perple_X (Connolly 1990) respectively. Both models predict that $Fe^{3+}$ is more compatible in pyroxenes compared to the observations from partitioning experiments. Note that the Jennings and Holland (2015) model predicts that $D_{Fe^{3+}}^{opx/melt}$ is greater than that of cpx/melt, which is opposite what is observed in experiments and xenoliths samples. For more details about the calculations, see text

**Table 1. Composition of starting mixes**

| Oxides | SM#3 | SM#6 | SM#7 |
|---|---|---|---|
| $SiO_2$ | 56.03(0.11) | 60.97(0.12) | 56.28(0.08) |
| $TiO_2$ | 0.44(0.0) | 1.11(0.02) | 0.45(0.03) |
| $Al_2O_3$ | 9.16(0.08) | 9.71(0.06) | 8.98(0.08) |
| $Cr_2O_3$ | - | - | 1.21(0.10) |
| FeO* | 6.56(0.08) | 10.25(0.07) | 6.75(0.08) |
| MgO | 10.10(0.05) | 8.87(0.07) | 9.89(0.06) |
| CaO | 13.35(0.05) | 4.68(0.08) | 13.17(0.04) |
| $Na_2O$ | 3.00(0.07) | 2.92(0.08) | 2.47(0.09) |
| $K_2O$ | 0.66(0.01) | 1.50(0.01) | 0.71(0.03) |
| Total | 99.30 | 100.01 | 99.91 |

numbers in the parentheses are 1σ standard deviations

**Table 2. Description of experiments**

| Experiment | Start. comp. | wire | initial T(°C) | final T(°C) | rate(°C/h) | phases[&] | ΔQFM |
|---|---|---|---|---|---|---|---|
| VF114 | SM#3 | Pt | 1251 | 1229 | 10 | cpx+gl | 0.55 |
| VF117 | SM#7 | Pt | 1258 | 1228 | 10 | cpx+gl | 0.86 |
| VF118 | SM#3 | Pt | 1251 | 1228 | 5 | cpx+gl | 1.63 |
| VF122 | SM#3 | Pt | 1252 | 1228 | 5 | cpx+gl | 0.08 |
| VF123[$] | SM#3 | Re | 1250 | 1229 | 5 | cpx+gl | -0.44 |
| VF133 | SM#6 | Pt | 1220 | 1207 | 5 | opx+gl | 0.57 |
| VF134[$] | SM#6 | Pt | 1220 | 1208 | 5 | opx+gl | 1.00 |
| VF139[$] | SM#6 | Re | 1220 | 1210 | 5 | opx+gl | -1.19 |
| VF140 | SM#6 | Pt | 1220 | 1208 | 5 | opx+gl | -0.57 |
| VF141 | SM#6 | Pt | 1220 | 1208 | 5 | opx+gl | 1.37 |
| VF147 | SM#3 | Pt | 1250 | 1230 | 5 | cpx+gl | 0.69 |
| VF150 | SM#3 | Pt | 1258 | 1232 | 5 | cpx+gl | 2.06 |

[$]Experiments which have been quenched by manual removal of the assembly

**Table 3. EPMA analyses of major element compositions**

| Expt. | $SiO_2$ | $TiO_2$ | $Al_2O_3$ | MgO | FeO* | $Na_2O$ | CaO | $K_2O$ | $Cr_2O_3$ | Total |
|---|---|---|---|---|---|---|---|---|---|---|
| Clinopyroxenes | | | | | | | | | | |
| VF114 | 54.03(0.13) | 0.14(0.02) | 0.98(0.08) | 19.16(0.16) | 3.43(0.15) | 0.18(0.02) | 21.05(0.29) | 0.01(0.01) | 0.03(0.03) | 99.10 |
| VF117 | 53.86(0.42) | 0.13(0.03) | 1.09(0.25) | 19.14(0.42) | 3.39(0.06) | 0.24(0.05) | 21.55(0.38) | 0.01(0.01) | 0.23(0.09) | 99.81 |
| VF118 | 54.80(0.39) | 0.10(0.03) | 0.64(0.16) | 20.00(0.32) | 3.13(0.18) | 0.17(0.01) | 21.61(0.45) | 0.01(0.01) | 0.04(0.02) | 100.74 |
| VF122 | 54.34(0.28) | 0.12(0.03) | 0.84(0.18) | 19.43(0.45) | 3.55(0.06) | 0.21(0.03) | 21.23(0.53) | 0.01(0.01) | 0.09(0.03) | 100.19 |
| VF123 | 53.50(0.41) | 0.11(0.01) | 0.56(0.10) | 20.33(0.25) | 3.84(0.11) | 0.16(0.01) | 20.98(0.35) | 0.01(0.01) | 0.05(0.01) | 99.53 |
| VF147 | 52.99(0.15) | 0.15(0.01) | 1.31(0.07) | 19.36(0.07) | 3.30(0.05) | 0.32(0.01) | 22.16(0.20) | - | 0.0490.02) | 99.94 |
| VF150 | 53.73(0.17) | 0.11(0.01) | 0.85(0.09) | 19.64(0.07) | 3.06(0.09) | 0.22(0.01) | 22.13(0.25) | - | 0.04(0.01) | 100.06 |
| Orthopyroxenes | | | | | | | | | | |
| VF133 | 55.86(0.01) | 0.16(0.01) | 0.43(0.02) | 31.87(0.08) | 10.08(0.04) | 0.03(0.01) | 0.90(0.03) | - | 0.05(0.02) | 99.38 |
| VF134 | 56.10(0.19) | 0.14(0.03) | 0.33(0.06) | 31.96(0.35) | 9.98(0.33) | 0.02(0.01) | 0.84(0.05) | 0.01(0.01) | 0.03(0.02) | 99.63 |
| VF139 | 55.09(0.27) | 0.17(0.02) | 0.36(0.04) | 30.01(0.57) | 12.56(0.61) | 0.02(0.01) | 1.02(0.11) | 0.01(0.00) | 0.04(0.01) | 99.55 |
| VF140 | 55.07(0.26) | 0.15(0.02) | 0.32(0.06) | 30.91(0.54) | 11.58(0.61) | 0.03(0.01) | 0.94(0.09) | 0.01(0.00) | 0.03(0.02) | 99.37 |
| VF141 | 55.77(0.28) | 0.14(0.02) | 0.39(0.06) | 33.67(0.68) | 8.74(0.75) | 0.01(0.00) | 0.41(0.06) | 0.01(0.01) | 0.03(0.01) | 99.46 |

Glasses

| Expt. | $SiO_2$ | $TiO_2$ | $Al_2O_3$ | MgO | FeO* | $Na_2O$ | CaO | $K_2O$ | $Cr_2O_3$ | Total |
|---|---|---|---|---|---|---|---|---|---|---|
| VF114 | 56.09(0.11) | 0.50(0.02) | 10.59(0.04) | 8.02(0.12) | 7.34(0.11) | 2.85(0.04) | 11.73(0.07) | 0.80(0.04) | - | 98.19 |
| VF117 | 57.16(0.41) | 0.50(0.02) | 10.72(0.19) | 8.17(0.14) | 6.82(0.05) | 2.97(0.04) | 11.80(0.33) | 0.79(0.02) | 0.01(0.01) | 99.18 |
| VF118 | 56.60(0.44) | 0.50(0.03) | 10.77(0.10) | 8.11(0.11) | 7.33(0.08) | 3.05(0.08) | 11.82(0.11) | 0.79(0.02) | - | 99.05 |
| VF122 | 56.95(0.15) | 0.52(0.01) | 10.98(0.07) | 8.20(0.02) | 7.05(0.08) | 3.36(0.03) | 12.02(0.05) | 0.79(0.01) | - | 100.07 |
| VF123 | 55.59(0.54) | 0.52(0.01) | 10.38(0.09) | 8.51(0.03) | 7.37(0.05) | 2.90(0.02) | 12.01(0.03) | 0.77(0.01) | 0.01(0.01) | 98.08 |
| VF147 | 56.21(0.11) | 0.51(0.01) | 10.48(0.06) | 8.50(0.03) | 7.14(0.09) | 3.44(0.09) | 12.66(0.01) | 0.02(0.00) | - | 99.00 |
| VF150 | 56.93(0.06) | 0.49(0.01) | 10.38(0.05) | 8.63(0.04) | 6.43(0.05) | 3.38(0.02) | 12.72(0.02) | 0.02(0.00) | - | 99.04 |
| VF133 | 60.96(0.17) | 1.17(0.02) | 11.12(0.10) | 6.72(0.05) | 9.60(0.07) | 3.75(0.11) | 5.09(0.01) | 1.58(0.01) | 0.01(0.01) | 100.02 |
| VF134 | 61.34(0.14) | 0.17(0.01) | 10.89(0.07) | 6.77(0.03) | 9.23(0.08) | 3.62(0.10) | 5.19(0.02) | 1.55(0.01) | - | 99.83 |
| VF139 | 61.22(0.10) | 1.19(0.01) | 11.43(0.07) | 5.71(0.03) | 9.73(0.08) | 3.59(0.12) | 5.24(0.02) | 1.41(0.41) | 0.01(0.01) | 99.56 |
| VF140 | 60.29(0.21) | 1.24(0.01) | 11.43(0.05) | 5.82(0.04) | 9.59(0.07) | 3.79(0.09) | 5.39(0.03) | 1.59(0.02) | 0.01(0.01) | 99.23 |
| VF141 | 60.87(0.09) | 1.20(0.01) | 11.19(0.04) | 6.07(0.03) | 9.12(0.08) | 3.99(0.09) | 5.39(0.02) | 1.58(0.01) | 0.00(0.01) | 99.57 |

**Table 4. XANES pre-edge centroid energies, $Fe^{3+}/Fe^{T}$ and calculated $D_{Fe^{3+}}$**

| | Glass | | clinopyroxene | | |
|---|---|---|---|---|---|
| Expt. | Centroid Energy[$] | $Fe^{3+}/Fe^{T}$[$] | Centroid Energy[$] | $Fe^{3+}/Fe^{T}$[$] | $DFe^{3+}$* |
| VF114 | 7112.468(0.022) | 0.199(0.015) | 7112.28(0.03) | 0.05(0.03) | 0.12(0.03) |
| VF117 | 7112.577(0.011) | 0.241(0.012) | 7112.40(0.06) | 0.12(0.05) | 0.24(0.03) |
| VF118 | 7112.723(0.017) | 0.309(0.012) | 7112.49(0.07) | 0.17(0.05) | 0.23(0.03) |
| VF122 | 7112.514(0.018) | 0.216(0.014) | 7112.22(0.04) | 0.01(0.04) | 0.03(0.03) |
| VF123 | 7112.322(0.014) | 0.151(0.016) | 7112.23(0.01) | 0.02(0.03) | 0.06(0.04) |
| VF147 | 7112.383(0.011) | 0.169(0.015) | 7112.46(0.13) | 0.15(0.08) | 0.41(0.07) |
| VF150 | 7112.637(0.016) | 0.268(0.012) | 7112.72(0.04) | 0.30(0.05) | 0.53(0.04) |
| | Glass | | orthopyroxene | | |
| VF133 | 7112.547(0.016) | 0.229(0.013) | 7112.23(0.03) | 0.06(0.02) | 0.27(0.04) |
| VF134 | 7112.417(0.019) | 0.181(0.015) | 7112.12(0.04) | 0.04(0.02) | 0.24(0.05) |
| VF139 | 7112.191(0.016) | 0.118(0.019) | 7112.04(0.02) | 0.03(0.02) | 0.29(0.11) |
| VF140 | 7112.228(0.038) | 0.126(0.020) | 7112.05(0.05) | 0.03(0.02) | 0.27(0.09) |
| VF141 | 7112.510(0.012) | 0.214(0.013) | 7112.15(0.03) | 0.05(0.02) | 0.21(0.05) |

*numbers in the parenthesis are standard errors

[$]numbers in the parenthesis are 1σ standard deviations

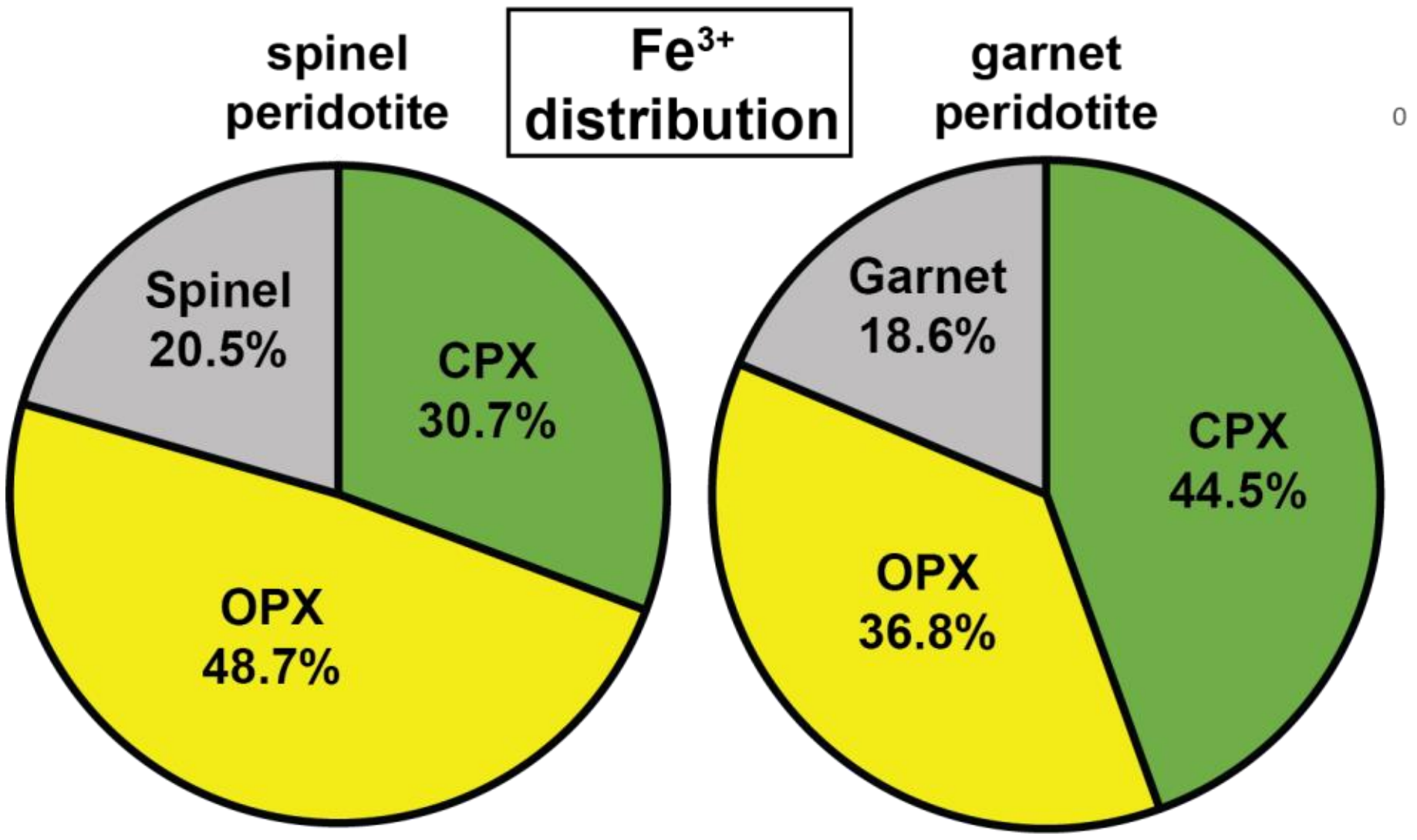

**Fig. 1**

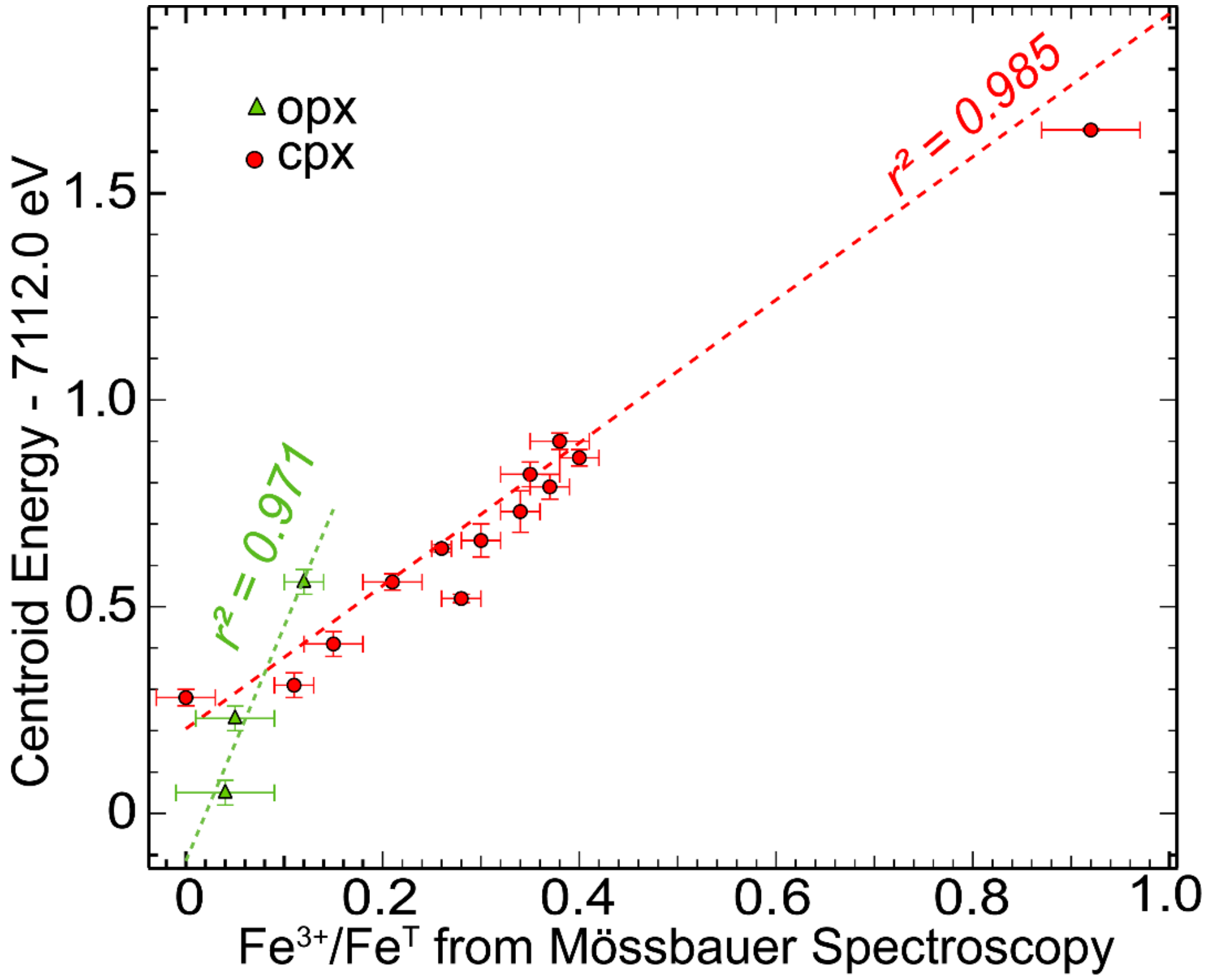

**Fig. 2**

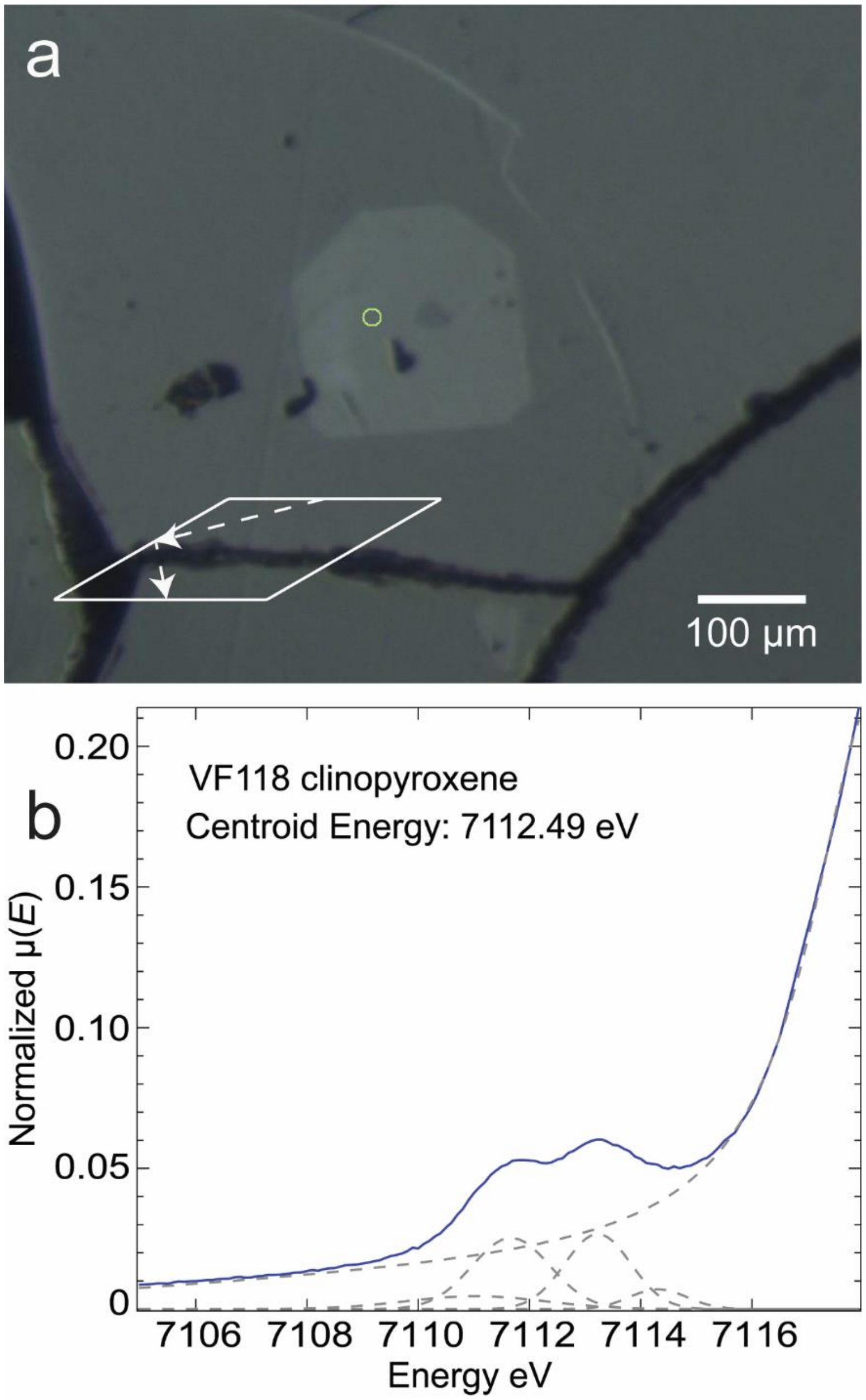

**Fig. 3**

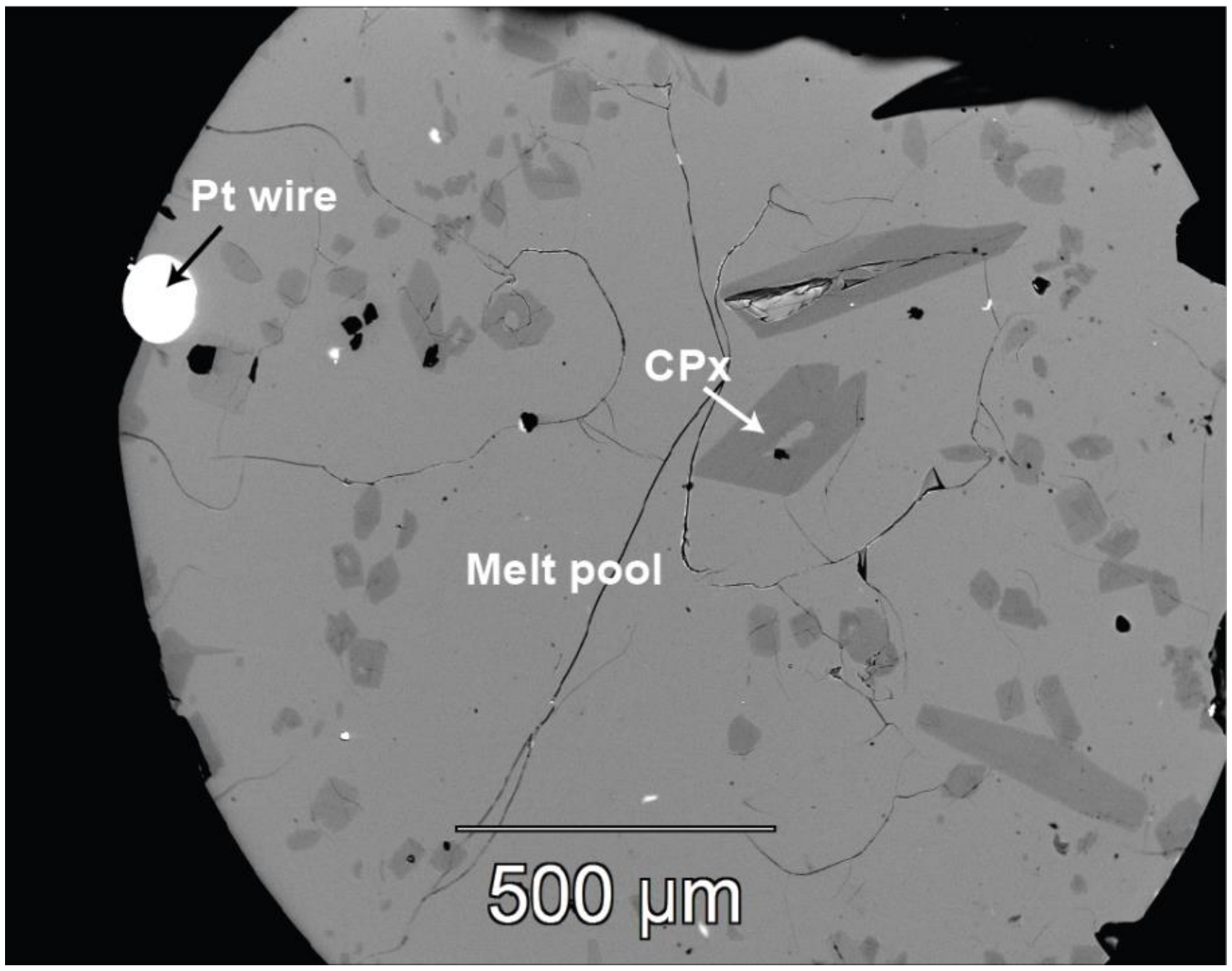

**Fig. 4**

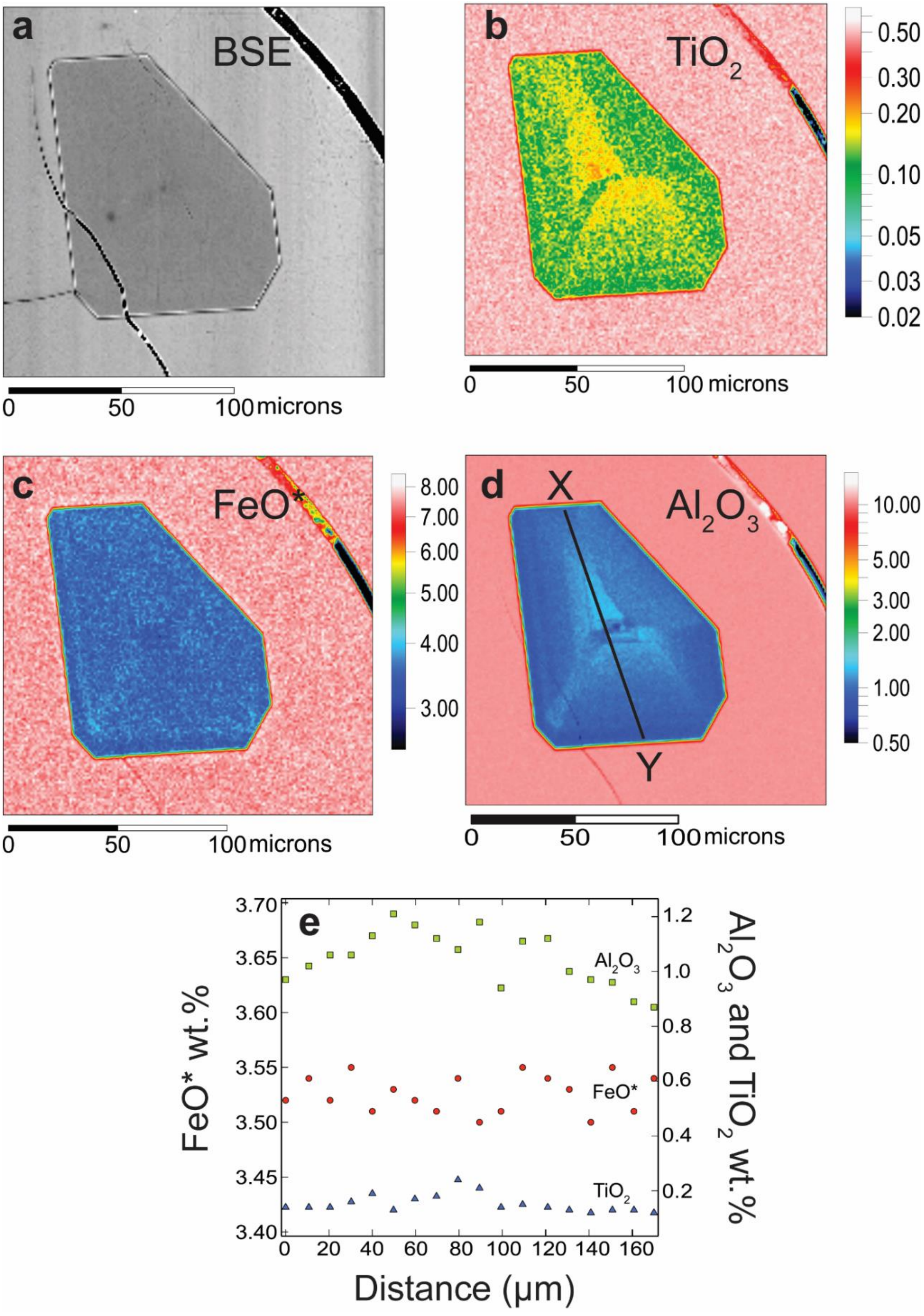

**Fig. 5**

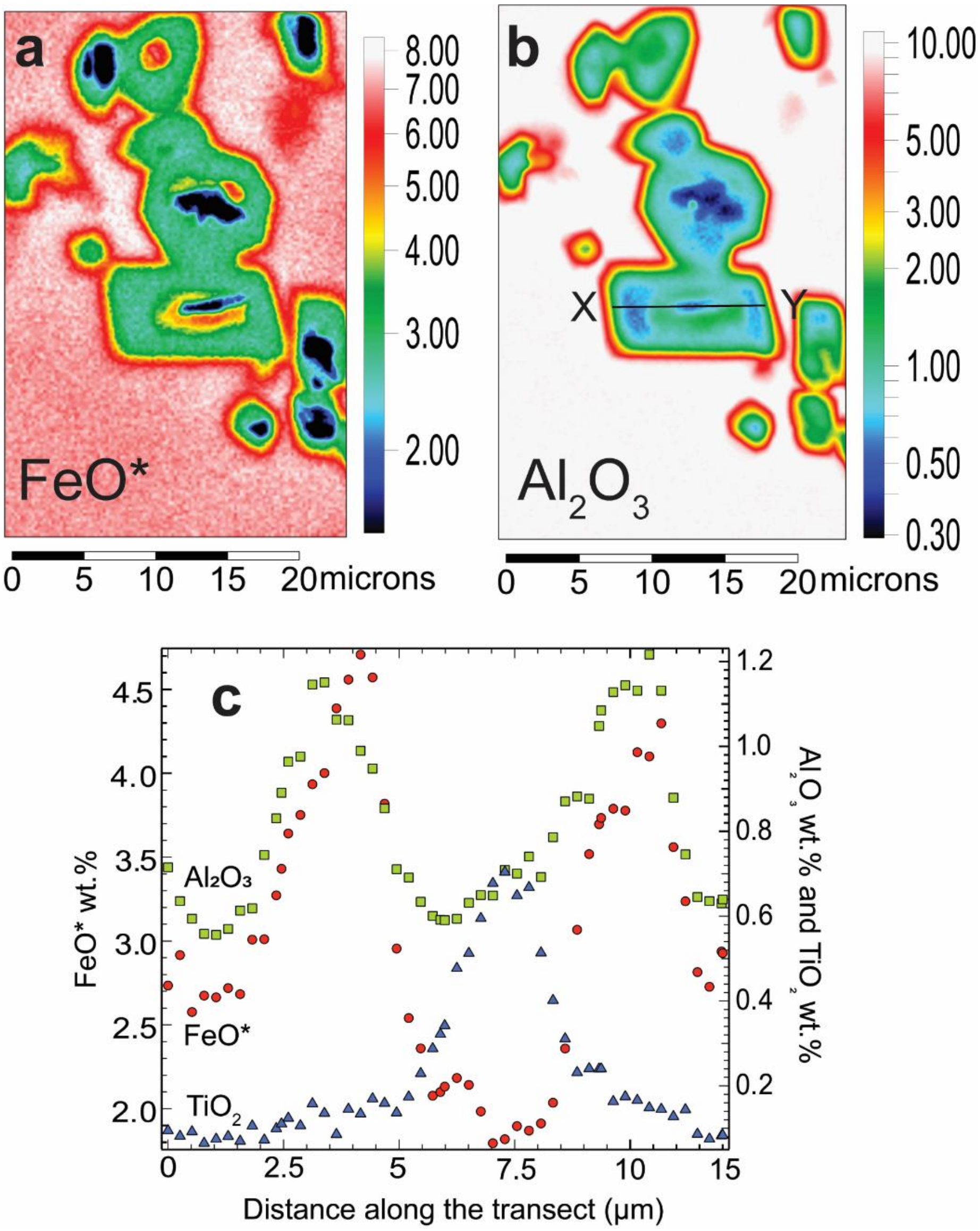

**Fig. 6**

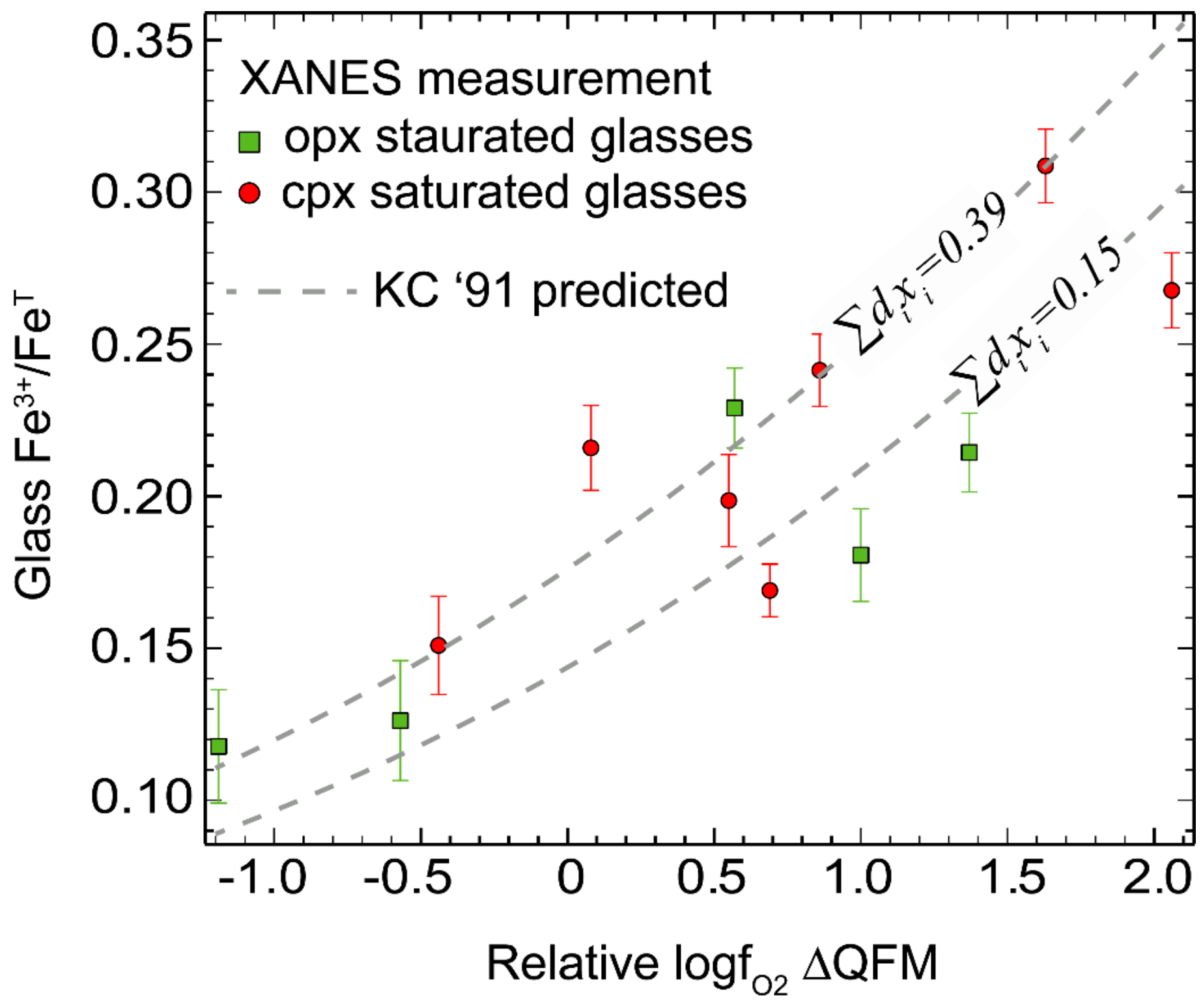

**Fig. 7**

.

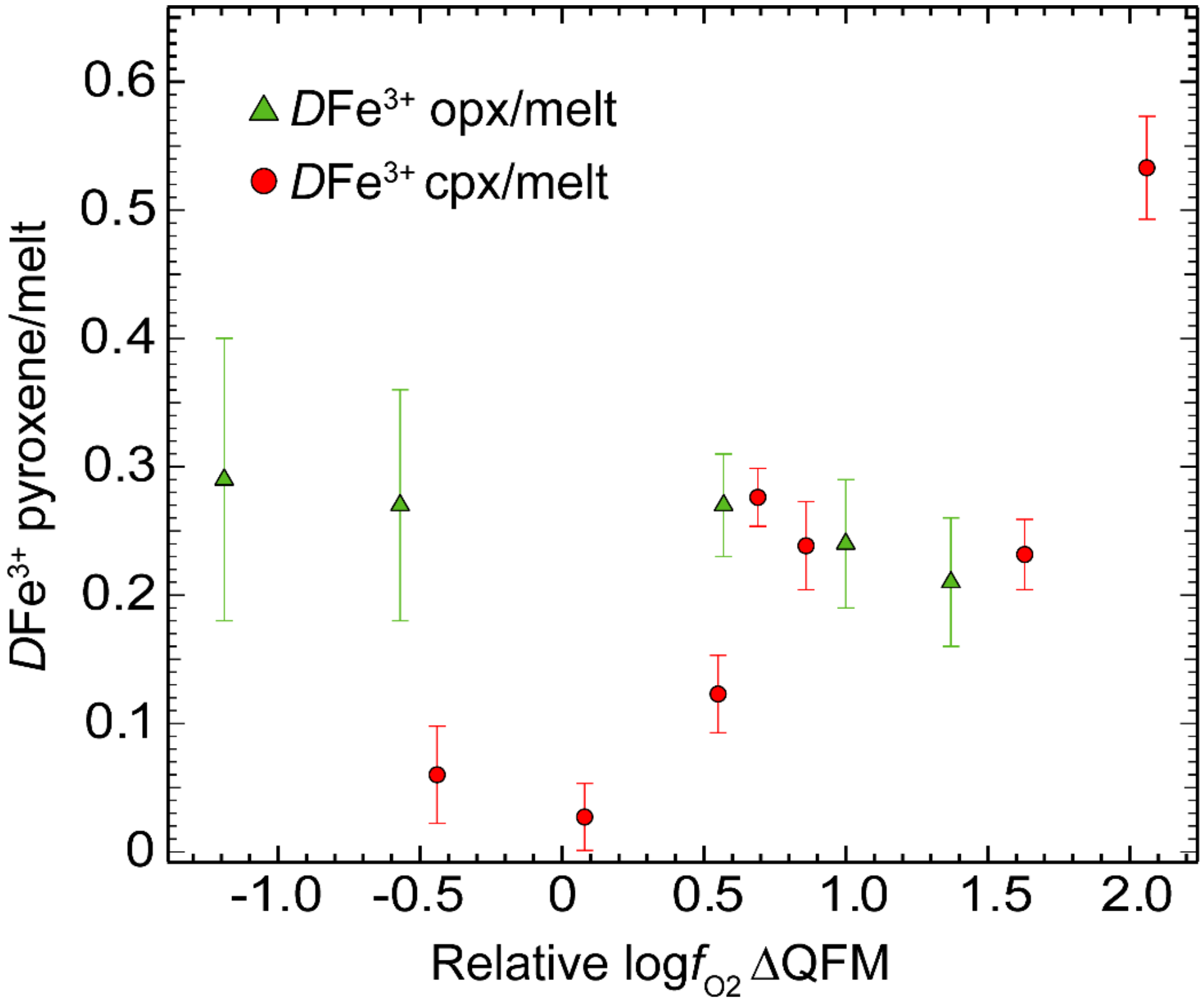

Fig. 8

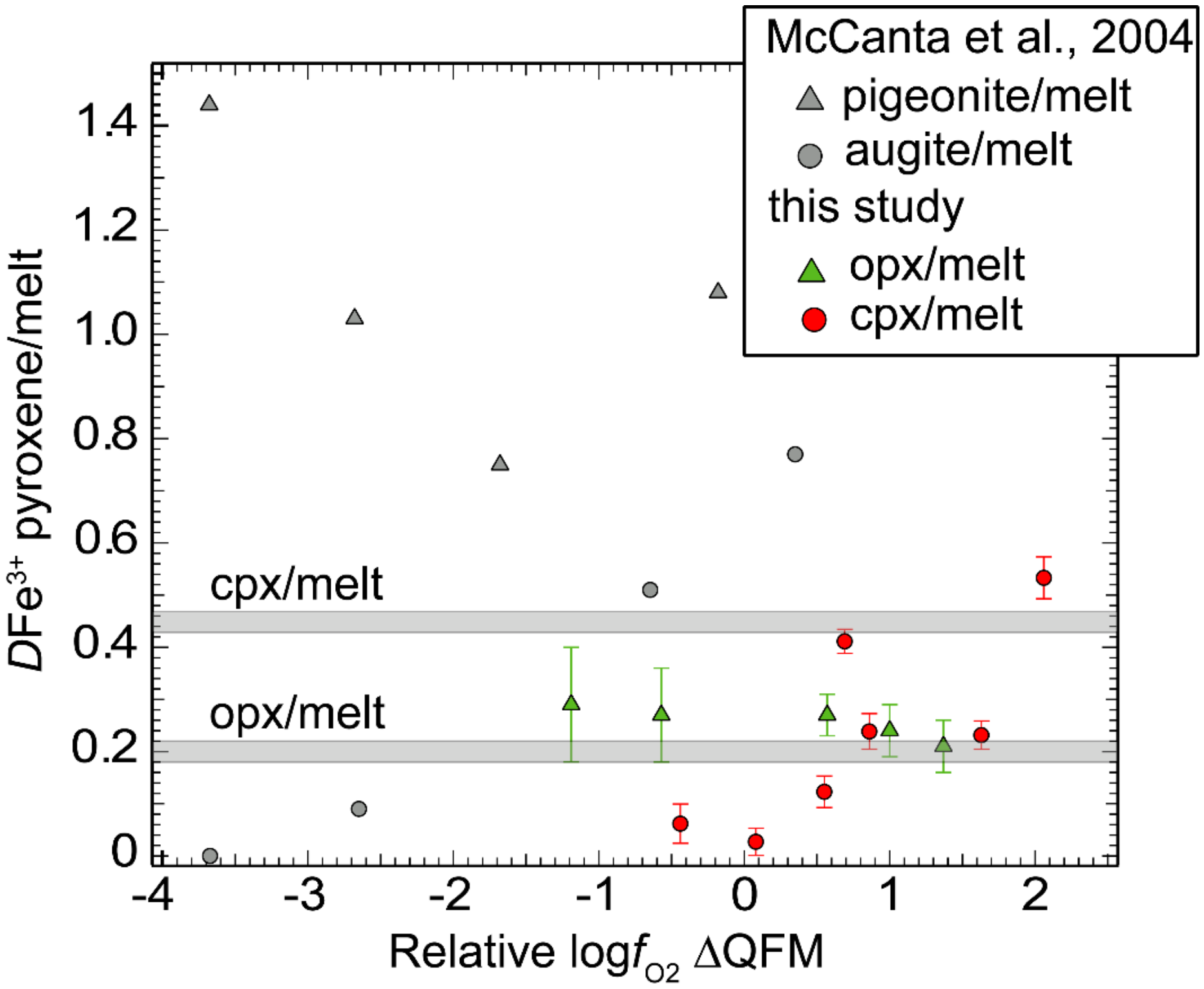

Fig. 9

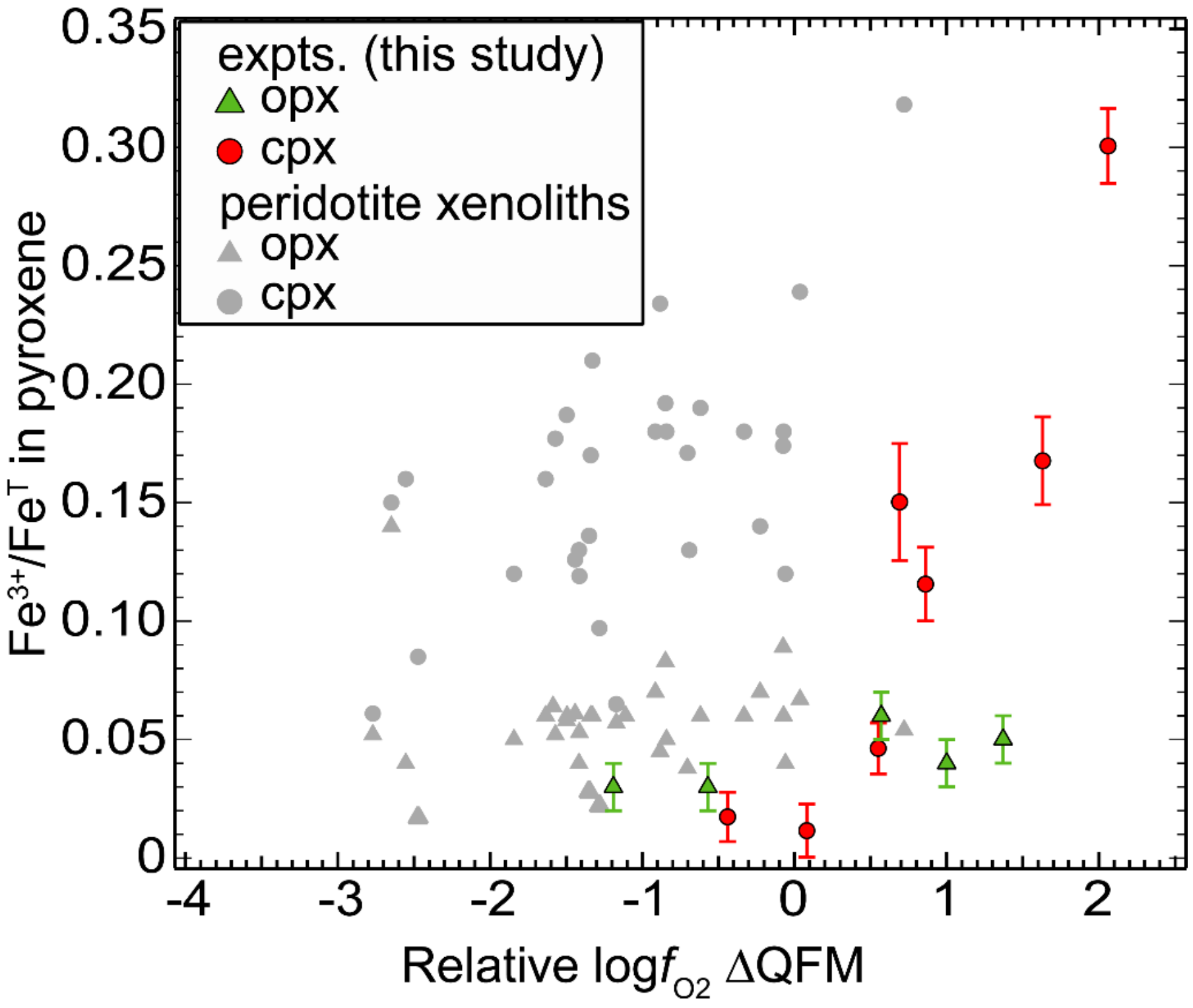

**Fig. 10**

.

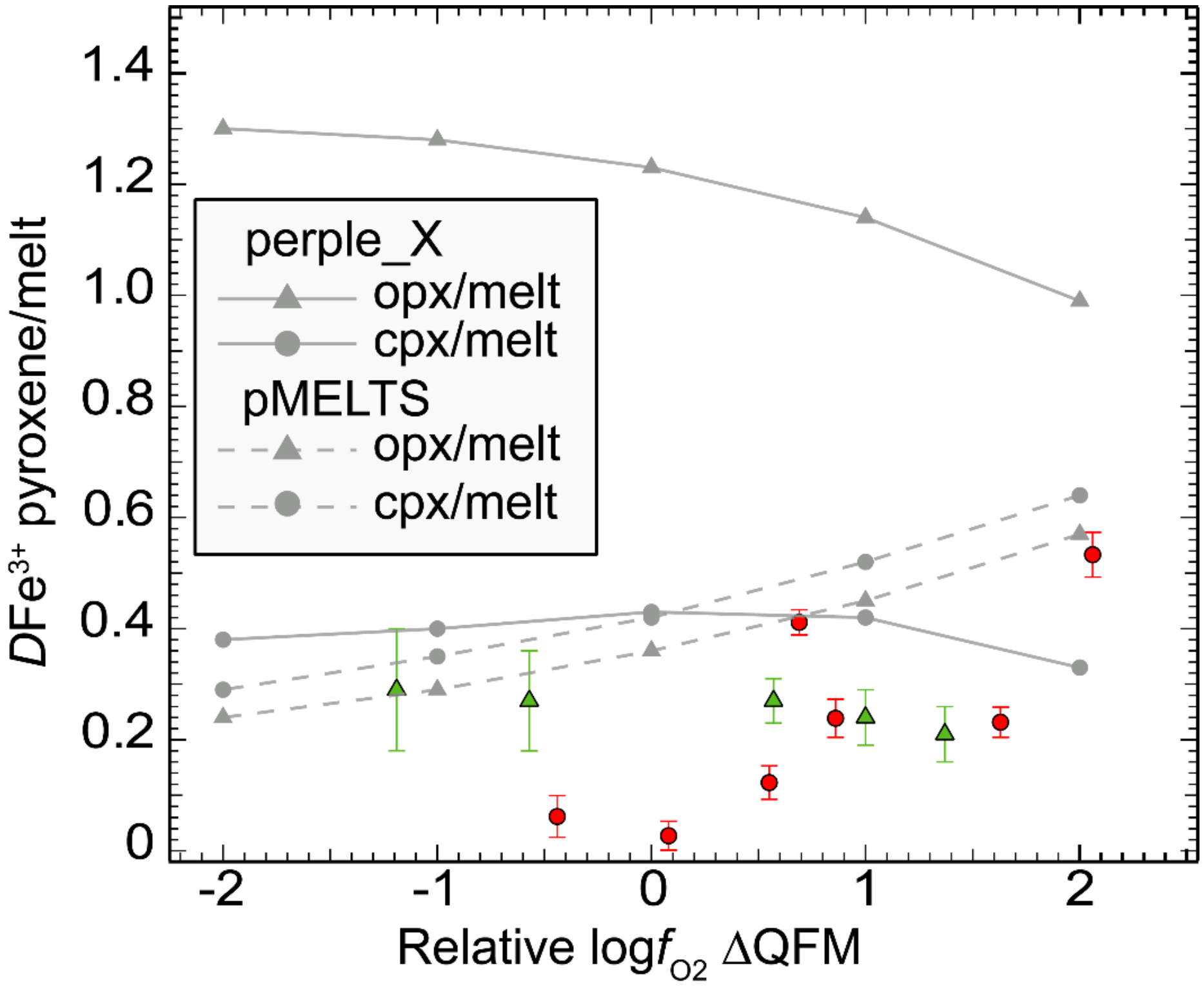

**Fig. 11**

**Supplementary Table 1.**

Description of pyroxene standards used in the calibration

| Sample name | Type | Mössbauer $Fe^{3+}/Fe^{T\$}$ | XANES centroid energy | Source | Ref. |
|---|---|---|---|---|---|
| 2015-039 | augite | 0.26(0.01) | 7112.641(0.009) | Canada Mus. Nat. | 1 |
| CMNMC43909 | hedenbergite | 0(0.03) | 7112.28(0.02) | Canada Mus. Nat. | 1 |
| 109098 | diopside | 0.37(0.02) | 7112.79(0.03) | Harvard Museum | 1 |
| MAL5 | augite | 0.34(0.02) | 7112.73(0.05) | Woodland, pers. collec. | 2 |
| MBR10 | augite | 0.28(0.02) | 7112.52(0.01) | Woodland, pers. collec. | 2 |
| MBR4 | augite | 0.38(0.03) | 7112.90(0.02) | Woodland, pers. collec. | 2 |
| MBR11 | augite | 0.35(0.03) | 7112.82(0.03) | Woodland, pers. collec. | 2 |
| MAL2 | augite | 0.30(0.02) | 7112.66(0.04) | Woodland, pers. collec. | 2 |
| MBR5 | augite | 0.40(0.04) | 7112.86(0.02) | Woodland, pers. collec. | 2 |
| NMNH176247 | aegirine | 0.92(0.05) | 7113.653(0.003) | Smithsonian Museum | 1 |
| JL8_cpx | Al-diopside | 0.11(0.02) | 7112.31(0.03) | Canil, pers. collec. | 3 |
| JL1_cpx | augite | 0.15(0.03) | 7112.41(0.03) | Canil, pers. collec. | 3 |
| RR222_cpx | Al-diopside | 0.21(0.03) | 7112.56(0.02) | Canil, pers. collec. | 3 |
| RR222_opx | enstatite | 0.05(0.03) | 7112.23(0.04) | Canil, pers. collec. | 3 |
| SL32_opx | enstatite | 0.12(0.03) | 7112.56(0.02) | Canil, pers. collec. | 3 |
| JL1_opx | enstatite | 0.04(0.03) | 7112.05(0.05) | Canil, pers. collec. | 3 |

$^{\$}$Mössbauer $Fe^{3+}/Fe^{T}$ has been corrected by a correction factor of *c*=1.2

Numbers in the parentheses are 1σ standard deviation.

Reference 1=this study; 2=Woodland and Jung 2007; 3= Luth and Canil 1993